\documentclass[aps,prd,showpacs,nofootinbib]{revtex4-2}

\usepackage{amsfonts}
\usepackage{pstricks}
\usepackage{xcolor}
\usepackage{amssymb}
\usepackage[tbtags]{amsmath}
\usepackage{mathtools}
\usepackage{bm}

\newcommand{\beq}{\begin{equation}}   
\newcommand{\eeq}{\end{equation}}

\usepackage{graphicx}
\usepackage{subfig}
\usepackage{caption}

\begin{document}



\title{Phase transitions and resilience of the MDCDW phase at finite temperature and density}


\author{William Gyory}
\affiliation{Department of Physics and Astronomy, University of Texas Rio Grande Valley}
\affiliation{Graduate Center, City University of New York}
\author{Vivian de la Incera}
\affiliation{Department of Physics and Astronomy, University of Texas Rio Grande Valley}
\affiliation{Graduate Center, City University of New York}
\begin{abstract}
\noindent

We study the phase transitions of the magnetic dual chiral density wave (MDCDW). This spatially inhomogeneous phase emerges in cold, dense QCD in the presence of a strong magnetic field. Starting from the generalized GL expansion of the free energy, we derive several analytical formulas that enable fast numerical computation of the expansion coefficients to arbitrary order, allowing high levels of precision in the determination of the physical dynamical parameters, as well as in the transition curves in the temperature vs. chemical potential plane at different magnetic fields. At magnetic fields and temperatures compatible with neutron star (NS) conditions, the MDCDW remains favored over the symmetric ground state at all densities. The phase's ``resilience" manifests in (1) a region of small but nonzero remnant mass and significant modulation at intermediate densities, originating in part from the nontrivial topology of the lowest Landau level, and (2) a region of increasing condensate parameters at high densities. Our analysis suggests the MDCDW condensate remains energetically favored at densities and temperatures much higher than previously considered, opening the possibility for this phase to be a viable candidate for the matter structure of even young neutron stars produced by NS mergers. 
\end{abstract}





\maketitle

\section{Introduction}
\noindent
In recent years, mapping the quantum chromodynamics (QCD) phase diagram has been a major goal of theoretical and experimental research. The most extreme temperatures and densities are relatively well understood because the QCD coupling constant becomes very small at high energy scales. This phenomenon, known as asymptotic freedom, enables rigorous calculations in the outermost regions of the phase diagram. The quark-gluon plasma (QGP) phase \cite{QGP} is predicted at high temperatures, and the color superconducting color-flavor-locked (CFL) phase \cite{alf-raj-wil-99/537, CS-Review} at low temperatures and asymptotically high densities. The regions of intermediate temperature and density remain much more difficult to investigate: The coupling constant is too large to allow for perturbative calculations, while the sign problem prevents the use of numerical calculations from lattice QCD. Exploring this region, therefore, requires using nonperturbative methods and effective theories.

Theoretical studies of the intermediate region have predicted a wide variety of quark matter phases that may be characterized by a ground state with either Cooper pairs or chiral condensates, depending on the quark chemical potential and temperature. Many of these phases have spatially inhomogeneous ground states \cite{CS-MRP86}-\cite{Nakano}. Therefore, it is a daunting task to determine the most energetically favored phase at any given region of the parameter space. Experimental probes could provide additional constraints on the theoretical work in this area; at the moment, however, there are no experiments capable of reaching the region of densities and temperatures required to confirm or rule out these models. Fortunately, upcoming heavy-ion collision (HIC) experiments, such as the RHIC Beam Energy Scan II (BES-II) \cite{Odyniec}, the Facility for Antiproton and Ion Research (FAIR) \cite{1607.01487} at the GSI site in Germany, and the Nuclotron-based Ion Collider Facility (NICA) \cite{Toneev} at JINR laboratory in Dubna, Russia, are all designed to run at unprecedented collision rates to provide high-precision measurements of observables in the higher baryon density and lower temperature region. 

The main tools available to constrain the theoretical models come from astrophysical observations of neutron stars (NSs), whose inner cores could reach densities high enough for quarks to deconfine. There are competing candidates for the matter in the core of an NS: It could be composed of neutrons, or, depending on the density, the neutrons could overlap to the point that the core matter becomes a ``soup" of quarks and gluons, forming one of the high-density/low-temperature phases predicted by theoreticians. Each phase will characteristically compress under gravity, leading to a different radius for a given mass and hence to a different equation of state (EoS). If astronomers can determine the EoS of an NS, that information could help pinpoint the phase of matter in the interior. It is far from trivial to obtain precise measurements of the radius and mass of NSs. However, much progress has recently been made thanks to new advanced instruments, such as the Neutron Star Interior Composition Explorer (NICER) telescope on the International Space Station \cite{MR-NICER} and the detection of NS mergers by the US Laser Interferometer Gravitational-Wave Observatory (LIGO) \cite{NSmerging}. 

Not only are NSs the natural objects where the intermediate phases of cold QCD could be realized, but they also exhibit strong magnetic fields, which become extremely large in the case of magnetars with observed surface fields $\sim10^{15}$ G. Magnetars can be produced after an NS merger via the instigation of various magnetic instabilities \cite{Science312}-\cite{AJ544}. The mechanisms associated with these magnetic instabilities during the merging process can lead to spinning magnetars with surface magnetic fields as large as $10^{17}$ G on a dynamical timescale, even if the two NSs that form the binary system have magnetic fields only of order $10^{11}-10^{13}$ G. The existence of this ultrastrong magnetic field is one of the most crucial factors for the realization of multimessenger astronomy. It is currently thought that the merger of the two NSs in the event GW170817 \cite{PRL119} produced such a magnetar, which was instrumental for the creation of the gamma-ray burst \cite{AJ848-L13} and the kilonova \cite{AJ848-L14,AJ856} that followed. 

Moreover, the magnetic fields of NSs may vary in strength from surface to the core. The scales of such variations, however, are much larger than the microscopic magnetic scale $\textit{l}_m$, which depends on the magnetic field strength. At both high and low fields, the star radius $R$ amply satisfies $R\gg \textit{l}_m$ \cite{Lattimer}; hence, in theoretical studies of the core matter phase, the magnetic field may be assumed to be constant and uniform. Theoretical calculations of the inner magnetic field strength based on the equipartition theorem give upper estimates on the order of $10^{18}$ G for nuclear matter \cite{Nuclear-matter-field} and $10^{20}$ G for quark matter \cite{Quark-Matter-field}. Because the vast majority of compact astrophysical objects have strong magnetic fields, and because these fields can significantly affect several properties of a star, many authors have been motivated to study the EoS of magnetized NSs \cite{Lattimer}-\cite{Aric}. Investigations of star stability for slowly rotating NSs allow for core fields up to just below $8 \times10^{18}(1.4M_{\odot}/M)$ G \cite{Cardall2001}. For rapidly rotating stars, a recent study \cite{2111.00013} based on magnetohydrodynamic simulations in full general relativity of self-consistent rotating NSs with ultrastrong mixed poloidal and toroidal magnetic fields found that poloidal field strengths in the star core can reach values a few times $\sim 10^{17}$ G. Therefore, it is reasonable to assume that magnetar inner fields can be in the range of $10^{17}-10^{18}$ G. 

Strong magnetic fields are also formed in off-central HIC. In the earliest moments after the collision, the system is subjected to a magnetic field of the order of $10^{18}$ G  \cite{NPA803}-\cite{AHEP2014}, the strongest ever created on Earth. Therefore, the two scenarios where quark deconfinement becomes relevant, NSs in nature and HIC experiments on Earth, typically also have very strong magnetic fields. This observation has motivated many studies on the effects of strong magnetic fields on quark matter phases.

From a fundamental point of view,  the presence of a magnetic field is also relevant due to the activation of new channels of interaction and, occasionally, also due to the generation of additional condensates. For instance, in the quarkyonic phase of dense quark matter, a magnetic field is responsible for the appearance of a new chiral spiral between the pion and magnetic moment condensates, $\langle\bar{\psi}\gamma^5\psi\rangle$ and $\langle \bar{\psi}\gamma^1\gamma^2\psi\rangle$ respectively \cite{Quarkyonic}.  Similarly, additional condensates emerge in the homogeneous chiral phase \cite{AMM-NJL}, as well as in color superconductivity \cite{CSB-6,CSB-7}.

Among the quark matter phases proposed at intermediate densities, phases with spatially inhomogeneous chiral condensates have long been viable candidates for the QCD phase map \cite{BubCar}. Investigations into Nambu-Jona-Lasinio (NJL)-like models \cite{NickelPRL, NickelPRD, Nakano}, quarkyonic matter \cite{Kojo}-\cite{ Kojo-2}, and the large-$N$ limit of QCD \cite{Rubakov}, for example, have featured single-modulated inhomogeneous chiral condensates under certain conditions. However, such phases in three spatial dimensions are subject to the Landau-Peierls (LP) instability \cite{LP}. The LP instability is characterized by the fact that thermal fluctuations of the Nambu-Goldstone bosons at nonzero temperatures, whose dispersions are anisotropic and soft in the direction normal to the modulation vector, wash out long-range order. However, some inhomogeneity remains due to algebraically decaying long-range correlations of the order parameter, forming a phase with a quasi--long-range order similar to liquid crystals. This effect has been shown to occur in the periodic real kink crystal \cite{HidakaPRD92}, the dual chiral density wave \cite{Lee-PRD92}, and quarkyonic matter \cite{PisarskiPRD99}. 

Although the LP instability effectively makes these inhomogeneous phases unstable at any finite temperature, a new outlook has emerged due to more recent studies that examine the effects of magnetic fields, thereby adding a third dimension to the phase diagram. Of particular interest to the present work is a dense quark matter phase in an external magnetic field that forms at intermediate densities when a dual chiral density wave ground state becomes energetically favored over the chirally symmetric one. This phase is known as the magnetic dual chiral density wave (MDCDW) phase \cite{Frolov}-\cite{NovelTop}. The MDCDW phase has profound differences from the so-called dual chiral density wave (DCDW) phase \cite{Nakano}, even though both are characterized by the same type of inhomogeneous chiral condensate $\langle\bar{\psi} \psi \rangle=\Delta \cos qz$, $\langle \bar{\psi} i\tau_3 \gamma_5 \psi \rangle=\Delta \sin qz$. 

In the absence of a magnetic field, the original symmetry of the two-flavor NJL model, where the DCDW phase is realized, is $SU_{V}(2)\times SU_{A}(2)\times SO(3)\times R^3$. In this case, the spontaneous breaking of chiral, rotational, and translational symmetries triggered by the inhomogeneous condensate gives rise to three independent Goldstone bosons, whose low-energy theory has soft transverse modes and hence exhibits the LP instability \cite{Lee-PRD92}. In comparison, when a magnetic field is present, it explicitly breaks the isospin and rotational symmetries, so that the original global symmetry of the two-flavor model is reduced to $U_{V}\times U_{A}\times SO(2)\times R^3$ \cite{NovelTop}. When the inhomogeneous chiral condensate forms, it breaks the chiral and translational symmetries, producing a single Goldstone boson. The low-energy theory of the fluctuations of this boson has no soft modes in the direction normal to the modulation vector. Thus the MDCDW phase is free of the LP instability \cite{AbsenceLP}. This result means that the MDCDW phase is not washed out by the fluctuations at low temperatures, in sharp contrast to inhomogeneous quark matter phases like the DCDW and others proposed for the core of neutron stars. Other important properties of the MDCDW phase that have no counterpart in other single-modulated phases are anomalous transport and the conversion of photons into axion polaritons in the MDCDW medium due to the anomalous coupling of photons with the axion field \cite{AP, MDCDW-Review}.

These discoveries warrant a systematic study into the behavior of the MDCDW phase in the regions relevant to potential physical applications. As mentioned above, for environments in which such extreme densities could be attained ---namely, neutron stars and future heavy-ion collision experiments---strong magnetic fields are expected to be present. While the effects of temperature and magnetic field on the phase with a DCDW condensate have been studied separately, there has been less research into the situation where both are present.  

Then, a natural and important task is to precisely calculate how the region in the $\mu$-$T$ plane where the MDCDW condensate is favored varies as the magnetic field strength is increased. As shown in this paper, the magnetic field extends the region of temperatures and densities where the inhomogeneous phase can exist to practically the entire region of parameters where the model is reliable. Notably, in the presence of a magnetic field, the inhomogeneous phase remains favored over the symmetric one at sufficiently low temperatures and intermediate to large baryon densities that range from 2.5 to about ten times the nuclear saturation density $n_0=0.16 \text{ fm}^{-3}$. When the chemical potential reaches values at which other inhomogeneous condensates would usually vanish, the magnitude of the MDCDW condensate remains small but nonzero, and the modulation continues to increase with the chemical potential $\mu$. With even larger $\mu$, the condensate magnitude starts growing again to sizable values. This resilience of the MDCDW ground state at low temperatures and intermediate to large densities has not been found in other inhomogeneous chiral phases. 

This paper explores the MDCDW phase at finite density and temperature using both a numerical minimization of the exact free energy and a generalized Ginzburg-Landau (GL) expansion. We derive an analytical expression that allows for fast computation of the expansion coefficients to arbitrary order, which is made possible using the Euler-Maclaurin formula. We then use the GL expansion to compute order parameters at various chemical potentials, temperatures, and magnetic fields. We also compare these results to those found by exact numerical minimization of the free energy to demonstrate the validity of the expansion. Finally, we use the GL expansion to generate phase diagrams showing how the magnetic field extends the region of the $\mu$-$T$ plane in which the inhomogeneous condensate is preferred; this effect becomes very significant at field strengths of order $10^{18}$ G, and remain noticeable even for fields of order $10^{17}$ G and temperatures commensurate with NS temperatures. These findings may thus prove relevant to the analysis of matter in neutron star interiors and future heavy-ion collisions, both of which are expected to contain magnetic fields of these orders.

The paper is organized as follows. In Sec. \ref{sec:GL Expansion}, we review the relevant NJL model and the generalized GL expansion method. We also present a set of formulas that allow for quick computation of every GL coefficient to arbitrary order. (We sketch derivations of these formulas in Appendices \ref{app:coef prefactors} \& \ref{app:alpha}.) We also show how the expansion in powers of the modulation $b$ effectively becomes an expansion in powers of $b/\mu<1$; hence, the expansion is valid in the region of interest, even though the order parameter $b$ is large. In Sec. \ref{sec:results}, we present several results displaying the phase diagrams of the MDCDW system in the $T$ vs $\mu$ plane. First, we compare the results obtained from numerical minimization of the exact free energy to those obtained from the GL expansion and comment on the accuracy of the approximation. Then we use the GL expansion (and one other technique, described in Appendix \ref{app:remnant T_c}) to determine the effects of magnetic field and temperature on the condensate and generate phase diagrams. In Sec. \ref{sec:discussion} we interpret our results from a physical perspective, explaining how the observed behavior relates to the pairing mechanisms driving the condensate. We also discuss how certain features of the solutions found for the dynamical parameters can be traced back to the nature of the GL coefficients and the expression for the free energy. We give concluding remarks in Sec. \ref{sec:conclusion}.

\section{Ginzburg-Landau Expansion Coefficients}
\label{sec:GL Expansion}

Let us consider a two-flavor effective theory of interacting quarks described by the following NJL model at finite baryon density in an external magnetic field, 
\begin{eqnarray} \label{L_NJL_QED}
\mathcal{L}=\bar{\psi}[i\gamma^{\mu}(\partial_\mu+iQA_{\mu})+\gamma_0 \mu]\psi 
+G[(\bar{\psi}\psi)^2+(\bar{\psi}i\tau\gamma_5\psi)^2].
\end{eqnarray}
Here, $Q=\mathrm{diag} (e_u,e_d)=\mathrm{diag} (\frac{2}{3}e,-\frac{1}{3}e)$, $\psi^T=(u,d)$, $\mu$ is the quark chemical potential, and $G$ is the four-fermion coupling. The electromagnetic potential $A^{\mu}=(0,0,Bx,0)$ corresponds to a constant and uniform magnetic field $\mathbf{B}$ pointing in the $z$-direction, with $x^{\mu}=(t,x,y,z)$. The Lagrangian (\ref{L_NJL_QED}) is symmetric under $U(1)_L\times U(1)_R \times SO(2)\times R^3$, reflecting the explicit breaking of flavor and rotational symmetries by the external field.  

To study the MDCDW phase, we introduce the following ansatz,
\begin{equation}
\label{eqn:condensate}
\langle\bar\psi\psi\rangle+i\langle\bar\psi i\gamma^5\tau_3\psi\rangle=\Delta e^{i q z}=-\frac{1}{2G}M(z). 
\end{equation}
This chiral density wave condensate is energetically favored over the homogeneous ones in a large region of chemical potentials \cite{Frolov, Tatsumi}.
 
The low-energy theory of the MDCDW phase can be explored using a generalized GL expansion. The thermodynamic potential is expanded in powers of the condensate $M(z)$ and its derivatives. Each term in this expansion must respect the symmetries of the two-flavor NJL model in a magnetic field. The MDCDW ansatz (\ref{eqn:condensate}) allows us to express the expansion in powers of $m=-2G\Delta$ and $b=q/2$, which are proportional to the condensate's magnitude and modulation, respectively. This procedure was carried out explicitly in \cite{AbsenceLP}, and it led to
\begin{align}\label{GL}
\Omega 	=\,\,	&\alpha_{2,0}m^2+\beta_{3,1}bm^2+\alpha_{4,0}m^4+\alpha_{4,2}b^2m^2+\beta_{5,1}bm^4\nonumber
\\
			&+\beta_{5,3}b^3m^2+\alpha_{6,0}m^6+\alpha_{6,2} b^2m^4+\alpha_{6,4}b^4m^2.
\end{align}

For convenience, and unlike \cite{AbsenceLP}, we wrote here the expansion in terms of $b=q/2$, instead of $q$, to remove needless factors of 2 from later formulas. Also, $b$ is more relevant than $q$ because $b/\mu<1$ in the region of interest: it will be seen below that the GL expansion at $B \neq 0$ effectively becomes an expansion in powers of $b/\mu$ after solving for the $\alpha$ and $\beta$ coefficients, as is the case when $B=0$ \cite{Carignano}. This feature is important because even though $b$ is large in the region of interest, the GL expansion remains reliable since $b/\mu$ is always small. We use $\alpha_{ij}$ and $\beta_{ij}$ (rather than $a_{ij}$ and $b_{ij}$) to distinguish the coefficients in (\ref{GL}) from those of \cite{AbsenceLP}.

The coefficients in the GL expansion are found from the derivatives of the thermodynamic potential. 
\begin{equation}
\label{eqn:free energy}
	\Omega	=\sum_f\left[ \Omega^f_{vac}(B)+\Omega^f_{anom}(B,\mu)+\Omega^f_\mu(B, \mu)+\Omega^f_T(B,\mu,T)\right]+\frac{m^2}{4G},
\end{equation}
where
\begingroup
\allowdisplaybreaks
\begin{align}
\label{eqn:free energyterms}		
	&\Omega^f_{vac}=\frac{1}{4\sqrt\pi}\frac{N_c\left|e_f B\right|}{(2\pi)^2}
			\int_{-\infty}^{+\infty}dk
			\sum_{\ell\xi\epsilon}
			\int_{1/\Lambda^2}^\infty\frac{ds}{s^{3/2}}e^{-s(E_\ell)^2}
		\\
	\label{anomfreeenergy}
	&\Omega^f_{anom}=-\frac{N_c\left|e_f B\right|}{(2\pi)^2}2b\mu
		\\
	\label{eqn:omega mu HLL}
	&\Omega^f_\mu=-\frac{N_c\left|e_f B\right|}{(2\pi)^2}\int_{-\infty}^{+\infty}dk
			\sum_{\xi,\ell>0}\left[(\mu-E_\ell)\theta(\mu-E_\ell)\right]\Big|_{\epsilon=+}+\Omega^{f,LLL}_\mu
		\\
	\label{eqn:omega T}
	&\Omega^f_T=-\frac{N_c\left|e_f B\right|}{(2\pi)^2}
		\frac{1}{\beta}
		\int_{-\infty}^{+\infty}dk
			\sum_{\ell\xi\epsilon}\ln{
				\left(
					1+e^{-\beta|E_\ell-\mu|}
				\right)}
		\\
	\label{eqn:omega mu LLL}
	&\Omega^{f,LLL}_\mu=-\frac{1}{2}\frac{N_c\left|e_f B\right|}{(2\pi)^2}
		\int_{-\infty}^{+\infty}dk
		\sum_\epsilon
			\left(
				|E_0-\mu|-|E_0|
			\right)_{reg},
\end{align}
\endgroup
and

\begin{equation}\label{LLLspectrum}
E_{0}=\epsilon\sqrt{m^2+k_3^2}+b,  \quad \epsilon=\pm,
\end{equation}  
\begin{equation}\label{HighLspectrum-1}
E_\ell= \epsilon\sqrt{(\xi\sqrt{m^2+k_3^2}+b)^2+2|e_fB|\ell}, \quad \epsilon=\pm, \xi=\pm, \ell=1,2,3,...
\end{equation}
are the quasiparticle energy modes for the lowest Landau level (LLL) ($\ell=0$) and for the higher Landau levels (HLL) ($\ell>0$) respectively.  For the HLL modes, $\epsilon$ indicates particle/antiparticle energies and $\xi$ the spin projection in the magnetic field direction. 

Notice that the LLL spectrum is asymmetric about zero energy. An asymmetric spectrum is a sign that the fermion structure possesses a nontrivial topology \cite{Niemi-Semenoff, Niemi}. This nontrivial topology is in turn reflected in various anomalous effects in the theory \cite{NovelTop, PLB69}, such as the anomalous term $\Omega^f_{anom}$, which is extracted after a careful regularization procedure based on an energy cutoff, as described in detail in \cite{Frolov}.  The same term can also be extracted by regularizing the Atiyah-Singer invariant \cite{Tatsumi}, which is a measure of the spectral asymmetry of the Hamiltonian. An explicit expression of the regularized $\Omega^{f,LLL}_\mu$ can be found in \cite{NovelTop}.
 
 The $\beta$ coefficients, which multiply odd powers of $b$, only get contributions from the LLL \cite{AbsenceLP} and hence have a topological origin. It is easy to see why the HLL does not contribute to the $\beta$ coefficients: If we change $b \to -b$ in the HLL terms of (\ref{eqn:free energyterms})--(\ref{eqn:omega mu LLL}), the HLL modes with $\xi=\pm1$ simply transform into each other, and the overall expression for $\Omega_{HLL}$ remains the same, meaning that $\Omega_{HLL}$ is even in $b$. This implies that the HLL terms can contribute to the $\alpha$ coefficients but not to the $\beta$ coefficients since the $\beta$ coefficients by definition multiply odd powers of $b$ in (\ref{GL}). For the LLL terms of (\ref{eqn:free energyterms})--(\ref{eqn:omega mu LLL}), there is no such invariance under $b \to -b$, so the LLL terms contribute to both even and odd powers of $b$ in the GL expansion and hence to both types of coefficients, $\alpha$ and $\beta$.  

 Solving for the coefficients in (\ref{GL}) is straightforward in principle, as it only requires taking derivatives of (\ref{eqn:free energyterms})--(\ref{eqn:omega mu LLL}) and then taking the limits $m,b\to 0$. In practice, however, this procedure runs into several challenges at the higher-order derivatives. First, the unwieldy form of (\ref{eqn:free energyterms})--(\ref{eqn:omega mu LLL}) makes taking high-order derivatives tedious. Second, as $|eB|$ decreases, the number of non-negligible Landau levels increases, so the sums in (\ref{eqn:free energyterms})--(\ref{eqn:omega mu LLL}) become computationally burdensome. Third, the integrand of the finite-temperature contribution $\Omega^f_T$ becomes increasingly oscillatory as more derivatives are taken. 

The above issues can be essentially eliminated by applying the Euler-Maclaurin formula to replace the Landau sums in (\ref{eqn:free energyterms})--(\ref{eqn:omega mu LLL}) with series in powers of $|eB|$. First, however, a few key observations and calculations help simplify the problem. First, it can be shown that all the coefficients of the same total order must differ only by numerical pre-factors. For example, $\alpha_{4,0}=\frac14\alpha_{4,2}$ and $\beta_{5,1}=\frac{3}{4}\beta_{5,3}.$ Denoting an arbitrary term in (\ref{GL}) by $c_{n,n_b}m^{n-n_b}b^{n_b}$, where $c=\alpha$ or $\beta$, we have the general formula
\begin{equation}
\label{eqn:coef prefactors}
	c_{n,n_b}=\frac{(n-2)!\,2^{1-(n-n_b)/2}}{(\frac{n-n_b}{2})!\,n_b!\,(n-n_b-2)!!}
	c_{n,n-2}.
\end{equation}
A proof of this formula is given in Appendix \ref{app:coef prefactors}. The task is thus reduced to computing only the coefficients of the form $\alpha_{n,n-2}$ and $\beta_{n,n-2}$.

The remaining task is to find all coefficients multiplying terms of the form $b^{n_b}m^2$. Beginning with the $\beta$ coefficients, note that these terms require taking only one derivative with respect to $m^2$ (and an odd number of derivatives with respect to $b$) and then taking the limits $m,b\to0$. It turns out that after taking the $m^2$-derivative, one $b$-derivative, and letting $m\to0$, we can arrive at a closed-form expression in terms of $b$ that allows for easy calculation of any number of remaining $b$-derivatives. Specifically, it can be shown that for $T>0$,
\begin{equation}
	\frac{\partial^2\Omega}{\partial b\partial(m^2)}\bigg|_{m=0}=\frac{3|eB|}{(2\pi)^2}
	\frac{1}{2\pi T}
	\operatorname{Re}\left[(-i)\,\psi^{(1)}\left(
		\frac{1}{2}+i\,\frac{\mu-b}{2\pi T}\right)\right]
		+f_{\text{odd}}(b),
\end{equation}
where $f_\text{odd}$ is an odd function. Taking $(n_b-1)$ more derivatives with respect to $b$ and then letting $b\to0,$ we find
\begin{equation}
\label{eqn:beta}
\beta_{n_b+2,n_b}=\frac{3|eB|}{(2\pi)^2}\cdot
\begin{dcases}
	\frac{1}{n_b!}\frac{1}{(2\pi T)^{n_b}}
		\operatorname{Re}\left[(-i)^{n_b}\psi^{(n_b)}
		\left(
			\frac{1}{2}+i\,\frac{\mu}{2\pi T}
		\right)\right] &\quad T>0, \\\\
	-\frac{1}{n_b\mu^{n_b}} &\quad T=0,
\end{dcases}
\end{equation}
where $n_b$ is odd and $\psi^{(n)}$ is the polygamma function of order $n$. 

From (\ref{eqn:beta}) it is apparent that the $\beta$ coefficients indeed vanish in the absence of a magnetic field, as expected. From (\ref{eqn:coef prefactors}) and (\ref{eqn:beta}) we can also see that at zero temperature, the general term $\beta_{n_m+n_b,n_b}m^{n_m}b^{n_b}$ becomes $C|eB|m^2(m/\mu)^{n_m-2}\allowbreak(b/\mu)^{n_b}$ for some dimensionless numerical factor $C$ after writing out $\beta$ explicitly. All powers of $b$ thus become powers of $b/\mu$, as claimed earlier. For $T>0,$ we can expand the polygamma function in (\ref{eqn:beta}) with a standard asymptotic series; this series adds correction terms to the $T=0$ case in powers of $T/\mu$, which is small and does not change the preceding argument.

We must still determine the $\alpha$ coefficients of the form $\alpha_{n_b+2,n_b}.$ As mentioned earlier, a direct calculation of these coefficients runs up against several technical issues, which are resolved using the Euler-Maclaurin formula for the HLL contributions. We can decompose each coefficient as $\alpha_{n_b+2,n_b}^{LLL}+\alpha_{n_b+2,n_b}^{HLL}$ and then apply the Euler-Maclaurin formula  to the HLL term, which gives a series in powers of $|eB|$. Only one term in this expansion has an odd power of $|eB|$, namely $|eB|^1$, and it turns out that this term exactly cancels the LLL contribution. Thus the final expression for each $\alpha$ coefficient takes the form of a series in even powers of $|eB|$. For $T>0$, we have  
\begingroup
\allowdisplaybreaks
\begin{align}
\label{eqn:alpha}
	\alpha_{n_b+2,n_b} &\sim\frac{\delta_{0,n_b}}{4G}
		+\sum_{j=0,2,4,\dots}|eB|^j\,\,\frac{B_j}{j!}\cdot
		\frac{1+2^j}{2\pi^2 3^{j-1}}\cdot
		\frac{1}{(n_b-1)!!}\,\,I_{n_b+2j-2}(\mu,T) &
		\nonumber\\\nonumber\\
	I_{-2}(\mu,T) &= 
		-\frac14\Lambda^2+\frac12\mu^2+\frac{\pi^2}3T^2
		\nonumber\\\nonumber\\
	I_0(\mu,T) &= 
		-\frac\gamma2-\left\{
			\ln\left(\frac{4\pi T}\Lambda\right)+\operatorname{Re}
			\left[
		        \psi\left(\frac12+i\frac\mu{2\pi T}\right)
			\right]
		\right\}
		\nonumber\\\nonumber\\
	I_{p>0}(\mu,T) &= 
		-\frac1p\left(\frac{i\sqrt2}\Lambda\right)^p
		-\frac1{p!!}
		\left\{\frac{1}{(2\pi T)^p}\operatorname{Re}
			\left[(-i)^p\psi^{(p)}\left(\frac12+i\,\frac\mu{2\pi T}\right)
			\right]
		\right\}.
\end{align}
\endgroup
For the $T=0$ case, we simply take the limit $T\to0$ on the terms in curly braces above, which are $\ln(2\mu/\Lambda)$ and $-(p-1)!\mu^{-p}$ respectively.
A sketch of this calculation is given in Appendix \ref{app:alpha}. Note that the general term $\alpha_{n_b+2,n_b}m^2b^{n_b}$ becomes a series with terms of the form $C\Lambda^2(|eB|/\mu^2)^jm^2\allowbreak(b/\mu)^{n_b}$ for some dimensionless factor $C$ (which may include powers of $\mu/\Lambda$ and $T/\mu$). So again, we see that powers of $b$ become powers of $b/\mu$ after solving for the coefficients, which supports the validity of the GL expansion in the region of interest.

Two features of formulas (\ref{eqn:coef prefactors})--(\ref{eqn:alpha}) are worth highlighting. First, these formulas allow for fast computation of any GL coefficient to arbitrary order. As mentioned above, the powers of $|eB|$ in (\ref{eqn:alpha}) effectively become powers of $|eB|/\mu^2$ after expanding the term $I_p(\mu,T)$. Even for large magnetic fields of order $10^{18}$ G, $|eB|/\mu^2$ is small in the region of interest, so the terms in the sum of (\ref{eqn:alpha}) decrease quickly. Moreover, the error term arising from the asymptotic nature of the Euler-Maclaurin formula remains small, so the coefficients given by (\ref{eqn:coef prefactors})--(\ref{eqn:alpha}) yield an accurate approximation of the true free energy of the system. This accuracy is discussed in Sec. \ref{sec:GL at B>0} and demonstrated in Fig. \ref{fig:OPs at B>0}, which compares the solutions for $m$ and $b$ obtained by minimizing the exact free energy with those obtained from the GL expansion.

Second, the expression for the $\alpha$ coefficients in (\ref{eqn:alpha}) remains valid in the limit $eB\to0$. (In fact, the error term arising from the Euler-Maclaurin formula vanishes in this case.) Every term in the series with $j>0$ vanishes, leaving only the $eB$-independent $j=0$ term. Thus, (\ref{eqn:coef prefactors})--(\ref{eqn:alpha}) give an exact closed-form expression for every GL coefficient in the zero--magnetic field (DCDW) case. Formula (\ref{eqn:beta}) also gives an exact closed-form expression for the $\beta$ coefficients since they depend only on the LLL and thus do not require the Euler-Maclaurin expansion. To our knowledge, no such formulas have been published, although particular low-order terms have been calculated. For example, Eq. (24) of [\citenum{Tatsumi}] gives an equivalent expression for $\beta_{3,1}$ (it differs by a factor of $-1/2$ only because of the way it is defined). An alternative formula for the $\alpha$ coefficients is given in Eq. (10) of [\citenum{NickelPRL}], but that expression involves both a sum and an integral.

\section{Results}
\label{sec:results}

\subsection{Validity of the GL expansion at $\bm{B=T=0}$}
\label{sec:GL at B=T=0}

To assess the validity of the GL expansion and coefficients given in (\ref{eqn:coef prefactors})--(\ref{eqn:alpha}), we have computed order parameters $m$ and $b$ using the exact free energy (\ref{eqn:free energy})--(\ref{eqn:omega mu LLL}) and also using the GL approximation at several different orders, similar to the approach taken in [\citenum{Carignano}]. Following [\citenum{Frolov}], we use proper-time regularization with $\Lambda\approx636.790$ MeV and coupling constant $G\Lambda^2=6$. We also work in the chiral limit, in which the quark current mass vanishes. These parameter values correspond to $m_{vac}=300$ MeV. All numerical computations involved only dimensionless quantities, e.g., $\tilde{m}=m/\Lambda$, and in both cases, we searched for global minima along a lattice in the $m$-$b$ plane with grid size $10^{-4}.$

\begin{figure}[h]
\centering
\includegraphics[width=.5\textwidth]{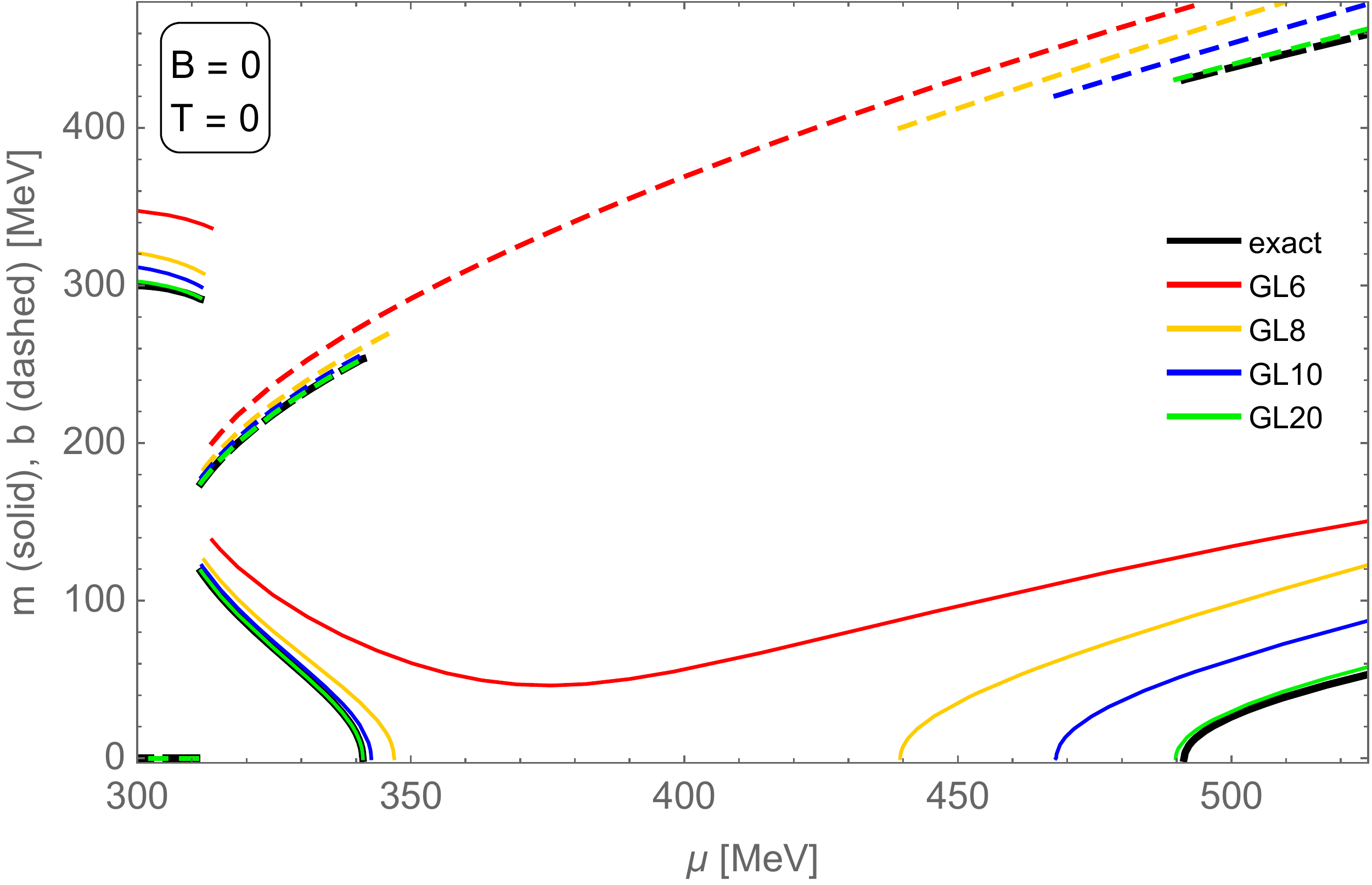}\hfill
\caption{Order parameters $m$ (solid) and $b$ (dashed) plotted against chemical potential at zero magnetic field and zero temperature. Black curves were computed by numerically minimizing the exact free energy, while colored curves were computed by minimizing the $n^\text{th}$-order GL approximation for $n=6,8,10,20$.}
\label{fig:OPs at B=T=0}
\end{figure}

Fig. \ref{fig:OPs at B=T=0} shows these results plotted against chemical potential in the simplest case, $B=T=0$. Recalling that formula (\ref{eqn:alpha}) gives the exact $\alpha$ coefficients in the limit $B\to0$, we expect the solutions calculated using the GL expansion to approximate the numerical solutions arbitrarily well at sufficiently high order, which is precisely what we see in Fig. \ref{fig:OPs at B=T=0}. In particular, the order parameters found using the $20^\text{th}$-order GL approximation are almost indistinguishable from the exact results. We wish to highlight that formulas (\ref{eqn:coef prefactors})--(\ref{eqn:alpha}) make it easy to calculate all coefficients up to $20^\text{th}$-order (or higher), enabling properties of the DCDW and MDCDW condensates to be computed much more quickly and easily than was previously possible.

Before proceeding to the $B>0$ and $T>0$ cases, let us highlight some key features of the phase behavior visible in Fig. \ref{fig:OPs at B=T=0}. The range of chemical potentials shown naturally separates into four distinct regions. In Region I (300–-311 MeV), there is no inhomogeneity ($b=0$), but there is a large nonzero $m\approx300$ MeV. After a first-order phase transition at $\mu=311$ MeV, spatial symmetry is broken by the existence of a nonzero inhomogeneity parameter $b$. In Region II (311--341 MeV), $m$ ($b$) smoothly decreases (increases) until $m$ vanishes in a second-order phase transition to Region III (341--491 MeV), in which chiral symmetry is restored, and the ground state is spatially homogeneous. Finally, in Region IV ($\mu>491$ MeV), the condensate returns, with $m$ taking on nonzero values and the inhomogeneity parameter $b$ attaining very large values. To our knowledge, the existence of Region IV, in which the condensate returns, has been overlooked (in the $B=0$ case) by previous authors. We discuss this region from a physical and mathematical perspective in Secs. \ref{sec:pairing} \& \ref{sec:GL analysis}.



\subsection{Validity of the GL expansion at $\bm{B>0}$ and $\bm{T>0}$}
\label{sec:GL at B>0}

\begin{figure}[h]
\centering
\subfloat{\includegraphics[width=.5\textwidth]{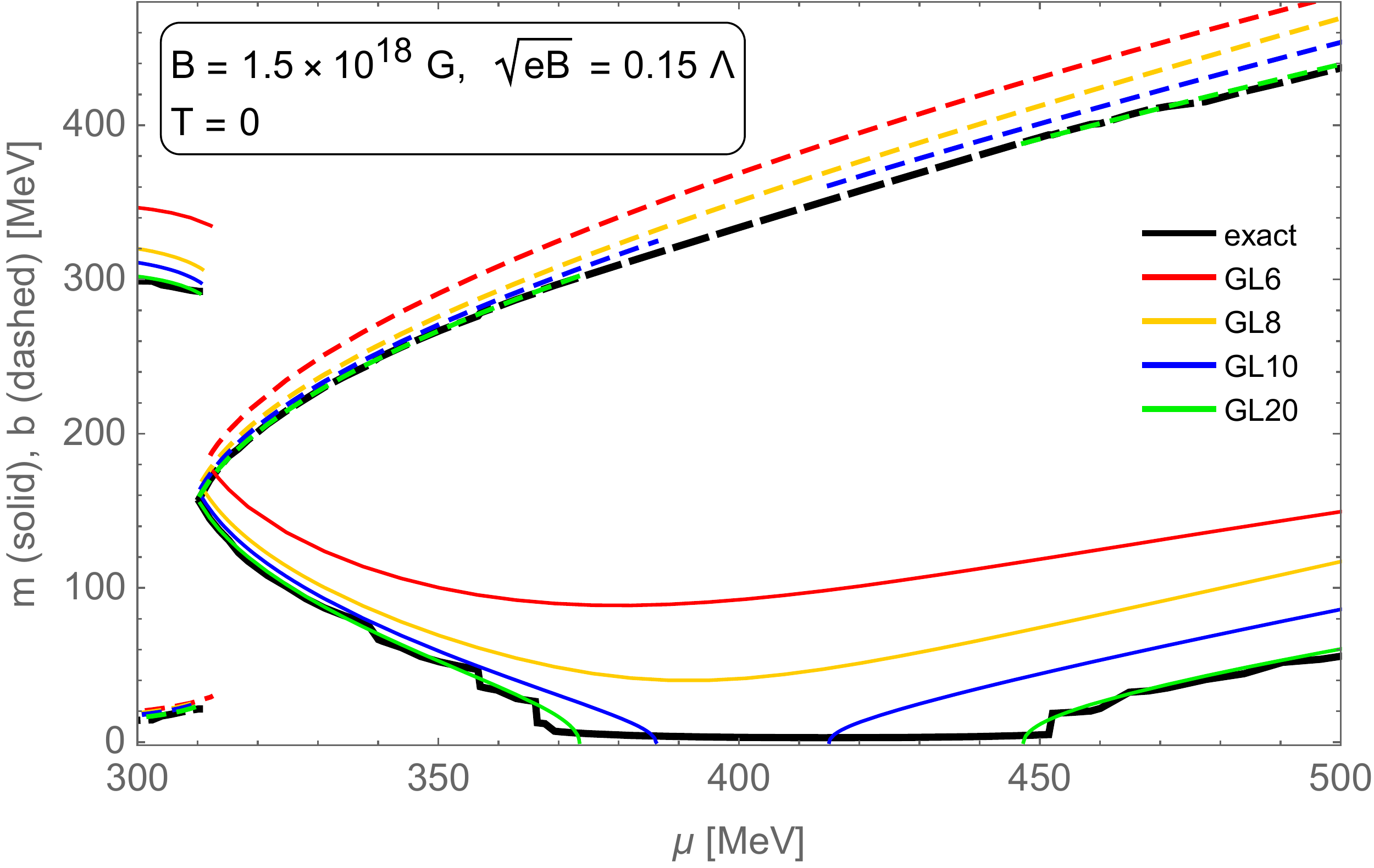}}\hfill
\subfloat{\includegraphics[width=.5\textwidth]{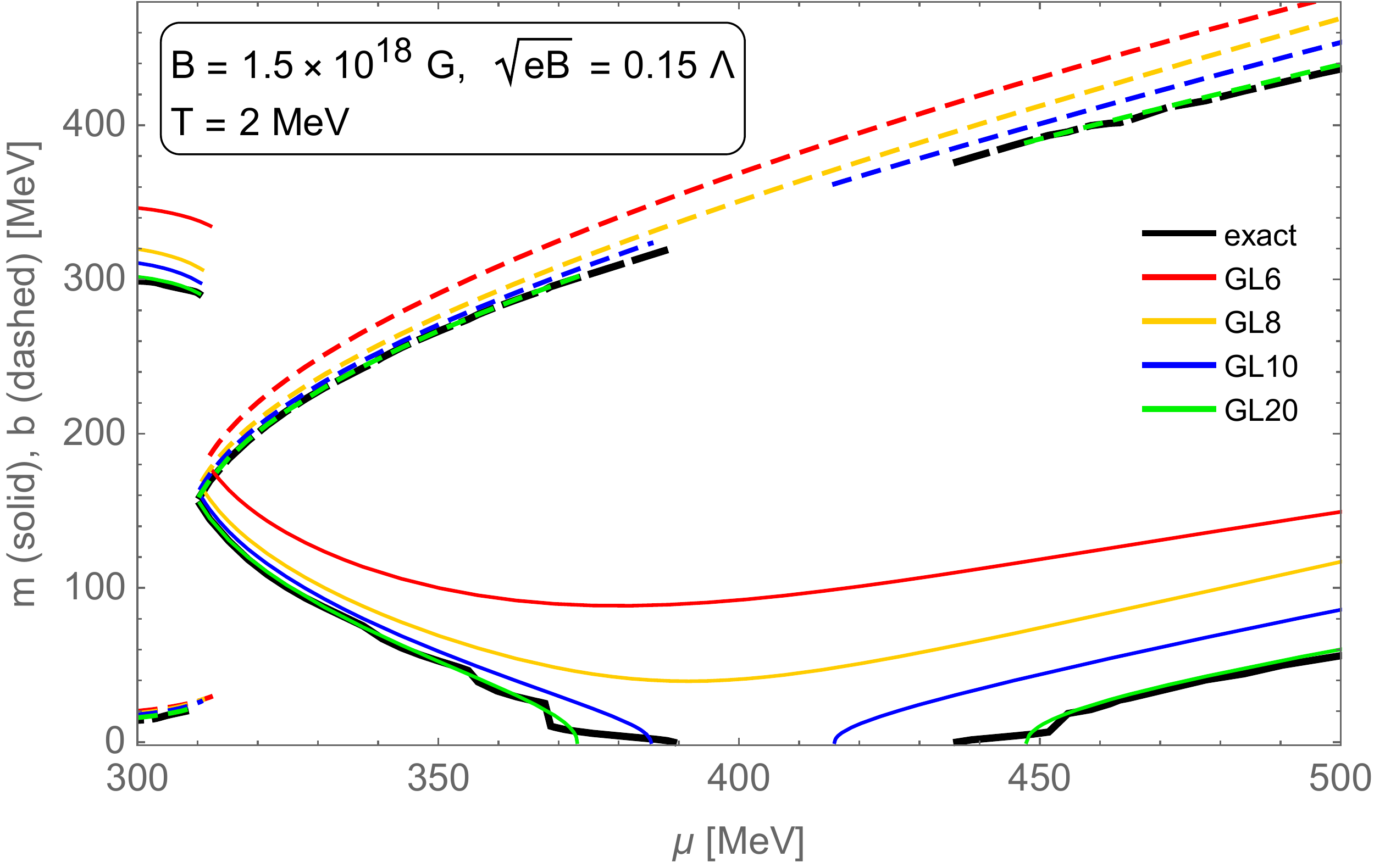}}\hfill
\\
\subfloat{\includegraphics[width=.5\textwidth]{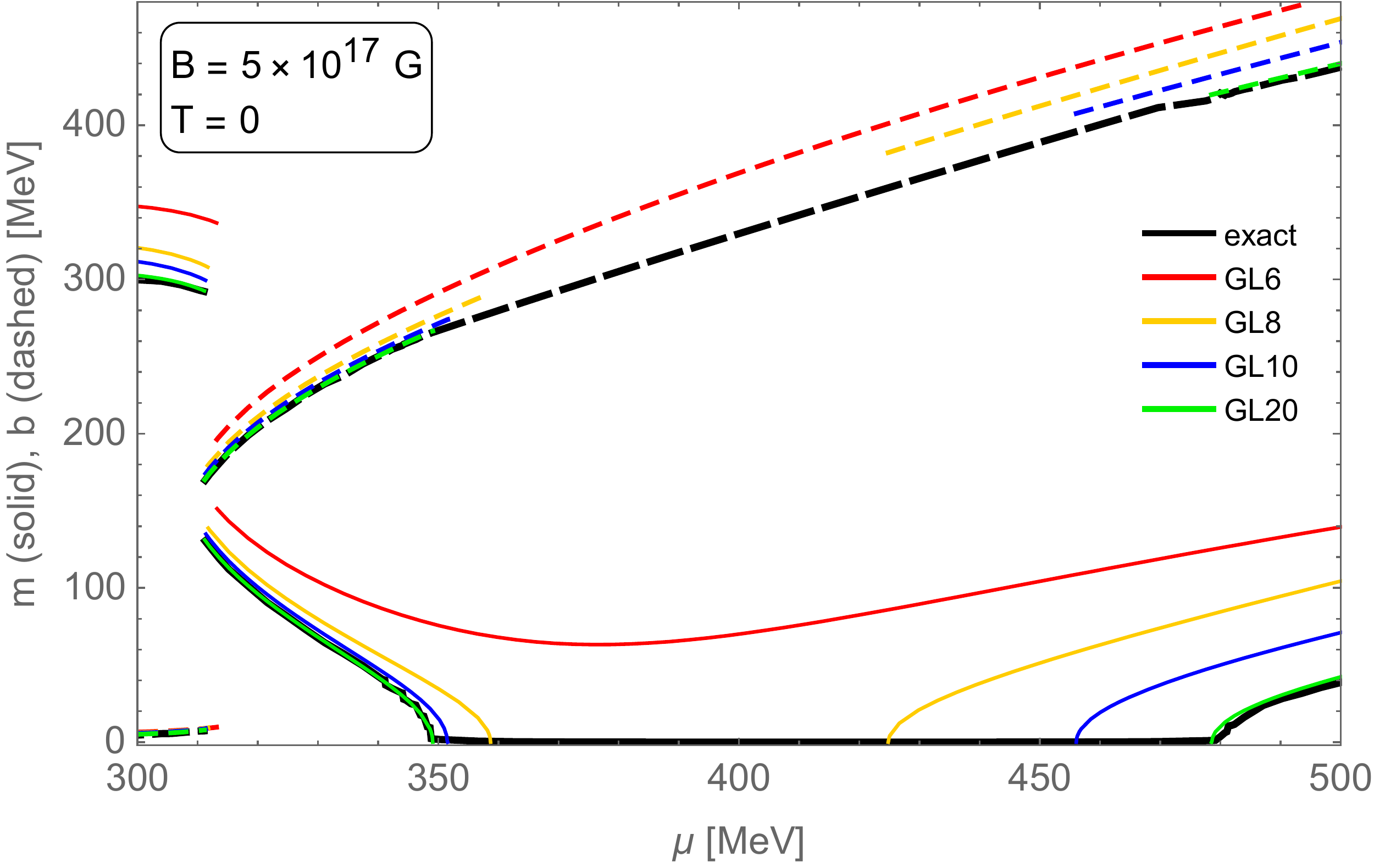}}\hfill
\subfloat{\includegraphics[width=.5\textwidth]{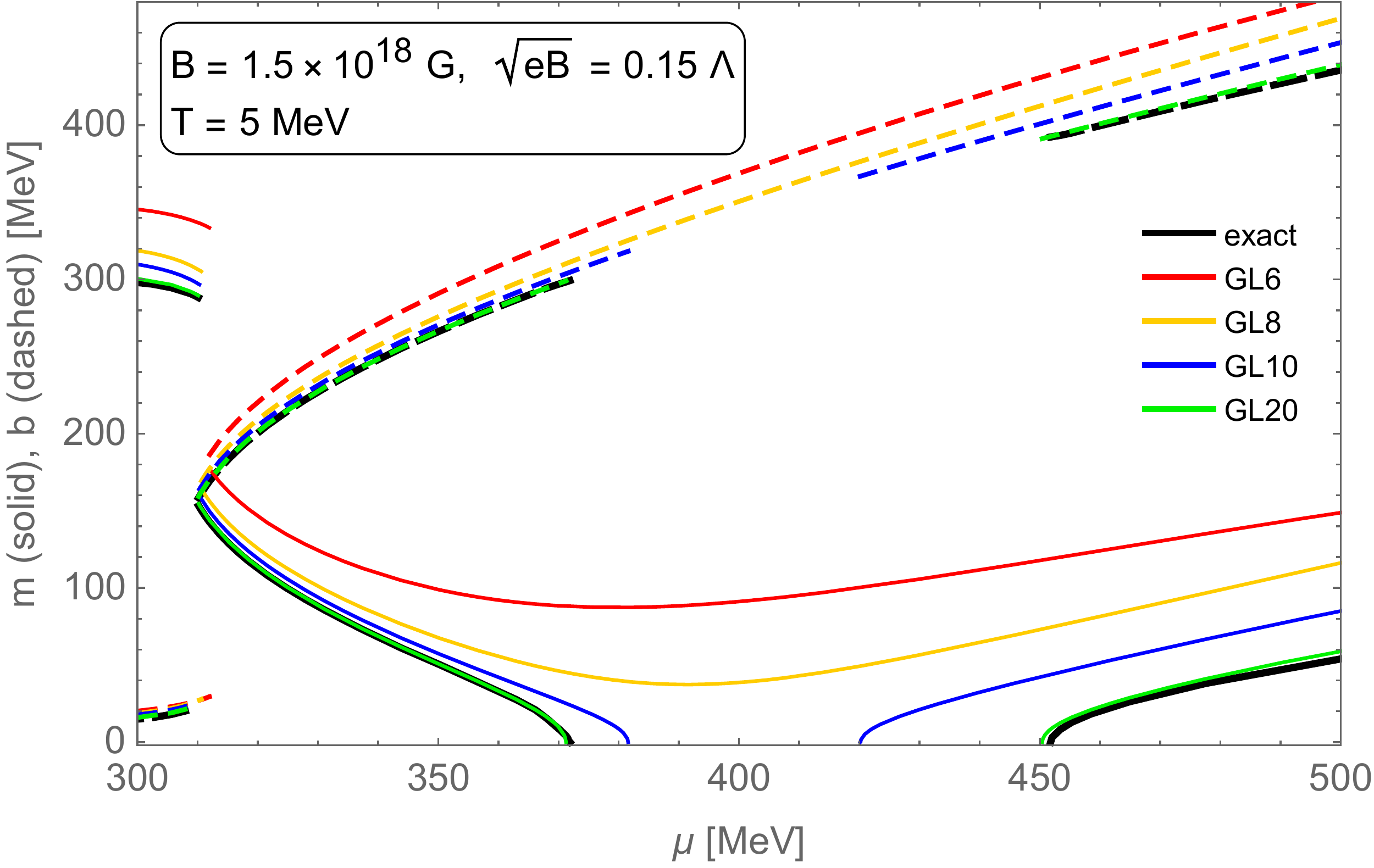}}\hfill
\caption{Comparison of order parameters computed from the exact free energy (black) and GL approximations of increasing order (colored) at various magnetic fields and temperatures. $m$ and $b$ are plotted with solid and dashed curves. \textbf{Left:} Magnetic field decreases from top to bottom while $T$ remains fixed at zero. The exact solutions at $T=0$ exhibit a region of small but nonzero ``remnant'' $m$ at intermediate $\mu$. One can distinguish between very small $m$ and $m=0$ by noticing that when $m$ is zero, there is no solution for $b$ because the potential becomes $b$-independent. The GL20 curve (green) provides an excellent, smoothed approximation of the exact curve, except in the region of the remnant mass, where it only captures it partially, as can be seen from the nonzero mass solutions in the two small segments at the beginning and the end of the remnant mass region found from the exact curve in the top plot. \textbf{Right:} Temperature increases from top to bottom while $B = 1.5\times10^{18}$ G remains fixed. Both plots can be compared to the top-left panel, in which $T=0$ for the same magnetic field. Increasing temperature smooths out the exact curve for $m$, erasing the remnant mass.}
\label{fig:OPs at B>0}
\end{figure}

The left panel of Fig. \ref{fig:OPs at B>0} shows the order parameters plotted for two field strengths, both at $T=0$. We highlight some fundamental differences in Regions I--IV compared to the $B=0$ case. Region I is now ``weakly inhomogeneous," as $b$ takes on a small nonzero value in this region. This inhomogeneity comes from the LLL anomalous contribution (\ref{anomfreeenergy}) to the free energy, which is a consequence of the system's nontrivial topology in the presence of a magnetic field, and hence is associated with the LLL spectral asymmetry. The anomalous term favors nonzero $b$, even when $\mu<m$;  higher Landau levels, on the other hand, contribute to Region I only via the vacuum part because the $\Theta$ functions in (\ref{eqn:free energy}) force the HLL medium term to be zero. We also discuss this phenomenon in Sec. \ref{sec:pairing}.

There is still a jump in the order parameters between Regions I and II. Still, it is no longer a phase transition because there is no change to any overall symmetry characterizing the state on either side. Compared to the $B=0$ case, Region II extends farther to the right, indicating that the magnetic field strengthens the inhomogeneous phase.  The exact (black) curve shows $m$ undergoing small jumps as it decreases. This non-smooth behavior results from the discretization of Landau levels in the exact numerical calculation, as opposed to the $B=0$ case that has a continuum of transverse momentum $k_\perp$. There is no such non-smooth behavior on the other colored curves at $B>0$ because they were found using the GL expansion with coefficients calculated using the Euler-Maclaurin formula. This formula approximates the Landau sums in (\ref{eqn:free energyterms})--(\ref{eqn:omega mu LLL}) with integrals; hence, the corresponding solutions provide a smoothed approximation to the exact solutions. Region IV shows a behavior similar to the zero-field case, although it covers a larger region of densities as it begins at smaller $\mu$.

Region III (368--452 MeV) is significantly different from that of the $B=0$ case: Instead of chiral symmetry restoration and homogeneity, we find that $m$ takes on a small but nonzero value, which we call the ``remnant mass." For $B=1.5\times10^{18}$ G, the remnant mass takes values of about 5 MeV at the outer edges of region III, reaching a minimum value of 3 MeV at the center. Because a global minimum of the free energy exists at $m>0$, $b$ remains well defined. As will be discussed in detail in Sec. \ref{sec:remnant mass} and Appendix \ref{app:remnant mass}, the origin of the remnant mass comes from two separate effects: the nontrivial topology of the LLL, which manifests in the presence of odd-in-$b$ terms in the GL expansion, and the behavior of the thermodynamic potential curvature at each Landau level $\ell>0$ along specific regions of the parameter space near $m=0$. The remnant mass is not fully found in the solutions obtained from the GL approximation because they were obtained using the Euler-Maclaurin formula, which approximates the Landau sums in $\ell>0$ by integrals. Notice that the order parameters computed with the $6^\text{th}$- and $8^\text{th}$-order GL expansions do not vanish over Region III in the top-left panel of Fig. \ref{fig:OPs at B>0} only because they are poor approximations, not because they are capturing the remnant mass effect. 

Observe that decreasing the magnetic field to $5\times10^{17}$ G brings the order parameters closer to their $B=0$ values, as expected (see Fig. \ref{fig:OPs at B>0}, bottom-left panel). However, as long as $T=0$ and $B>0$, the remnant mass cannot truly vanish (though it can become tiny). With the introduction of nonzero temperature, however, the remnant mass and the non-smooth behavior of $m$ are eventually eliminated (see Fig. \ref{fig:OPs at B>0}, right panels). In each panel of Fig. \ref{fig:OPs at B>0}, one can distinguish regions with very small $m$ from those where $m=0$ by realizing that there is a $b$ solution in the former case but never in the latter one, because the thermodynamic potential cannot depend on the modulation of a condensate whose magnitude is zero and hence does not exist. Once $m=0$, $b$ becomes undefined since $\Omega$ becomes $b$-independent. For example, compare the exact (black) curves for $m$ and $b$ in Region III of the left panels of Fig. \ref{fig:OPs at B>0} (where $T=0$ and hence $m$ takes on remnant values, so $b$ is nonzero) with those of the right panels (where $T>0$, and hence $m$ vanishes and $b$ is undefined over Region III). 

It is worth mentioning that a sufficiently high-order GL expansion with coefficients given by (\ref{eqn:coef prefactors})--(\ref{eqn:alpha}) always yields a good approximation to the order parameters in Regions I--IV, provided we are willing to neglect the remnant mass and accept a smoothed average of $m$. Moreover, when $T$ is increased to roughly the same order of magnitude as the remnant mass, the already-minor discrepancies between the exact and GL solutions become even smaller. Indeed, with the tools developed in this paper, one can easily increase the GL expansion's accuracy to any order. This and the discovery of the remnant mass are two main results of the present work. 

\subsection{Effects of temperature on the condensate}

\begin{figure}[h]
\centering
\subfloat{\includegraphics[width=.5\textwidth]{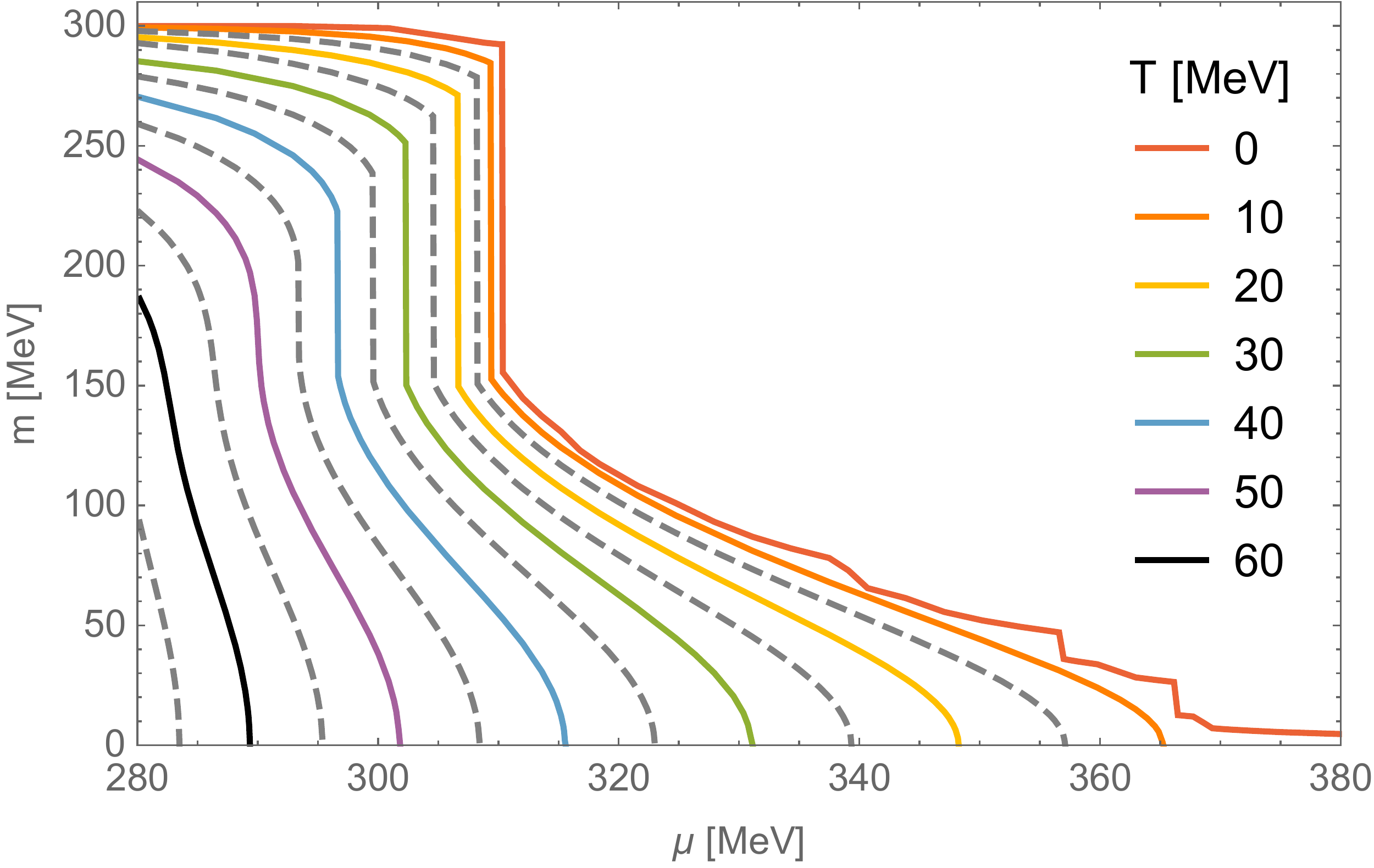}}\hfill
\subfloat{\includegraphics[width=.5\textwidth]{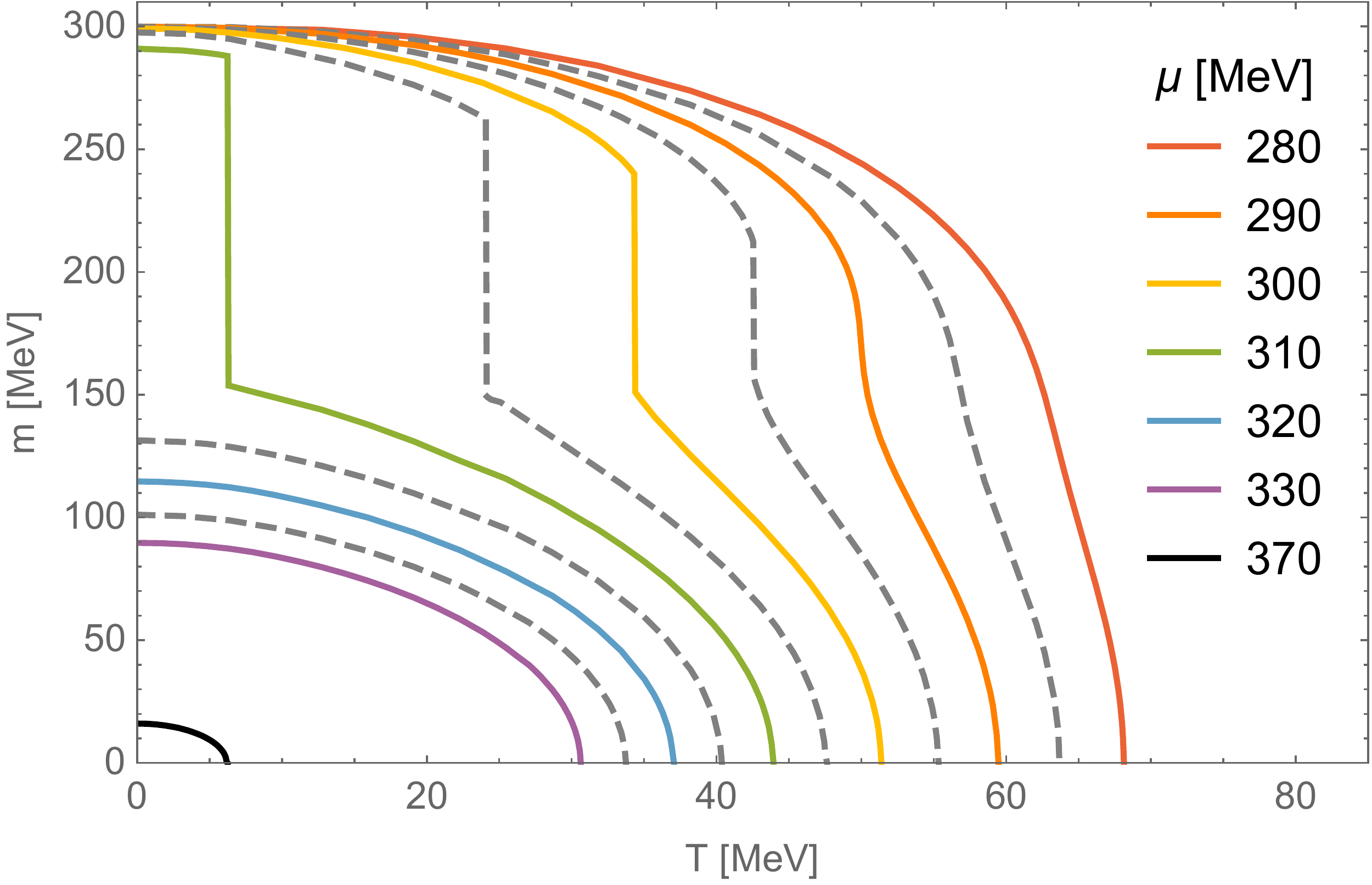}}\hfill
\\
\subfloat{\includegraphics[width=.5\textwidth]{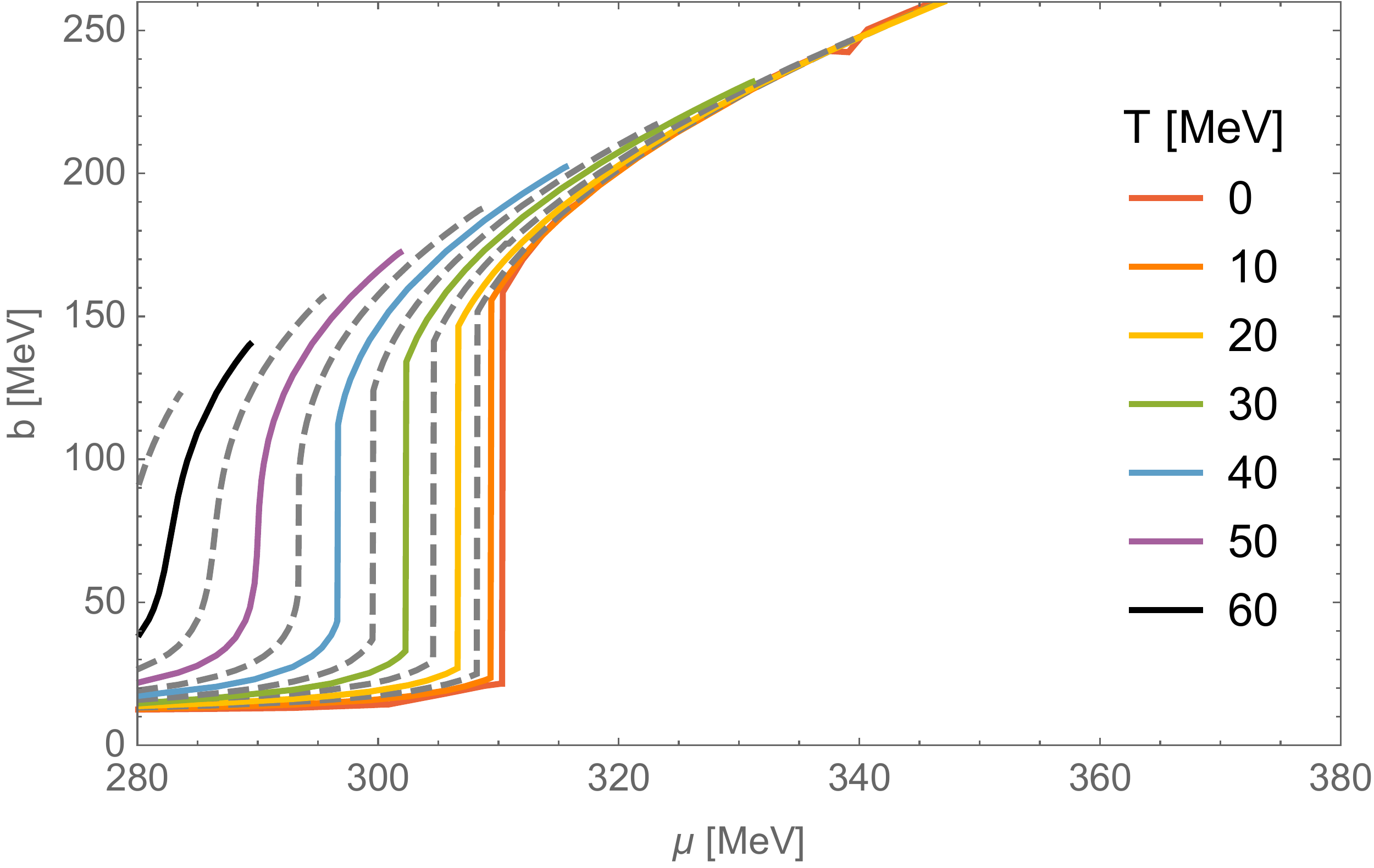}}\hfill
\subfloat{\includegraphics[width=.5\textwidth]{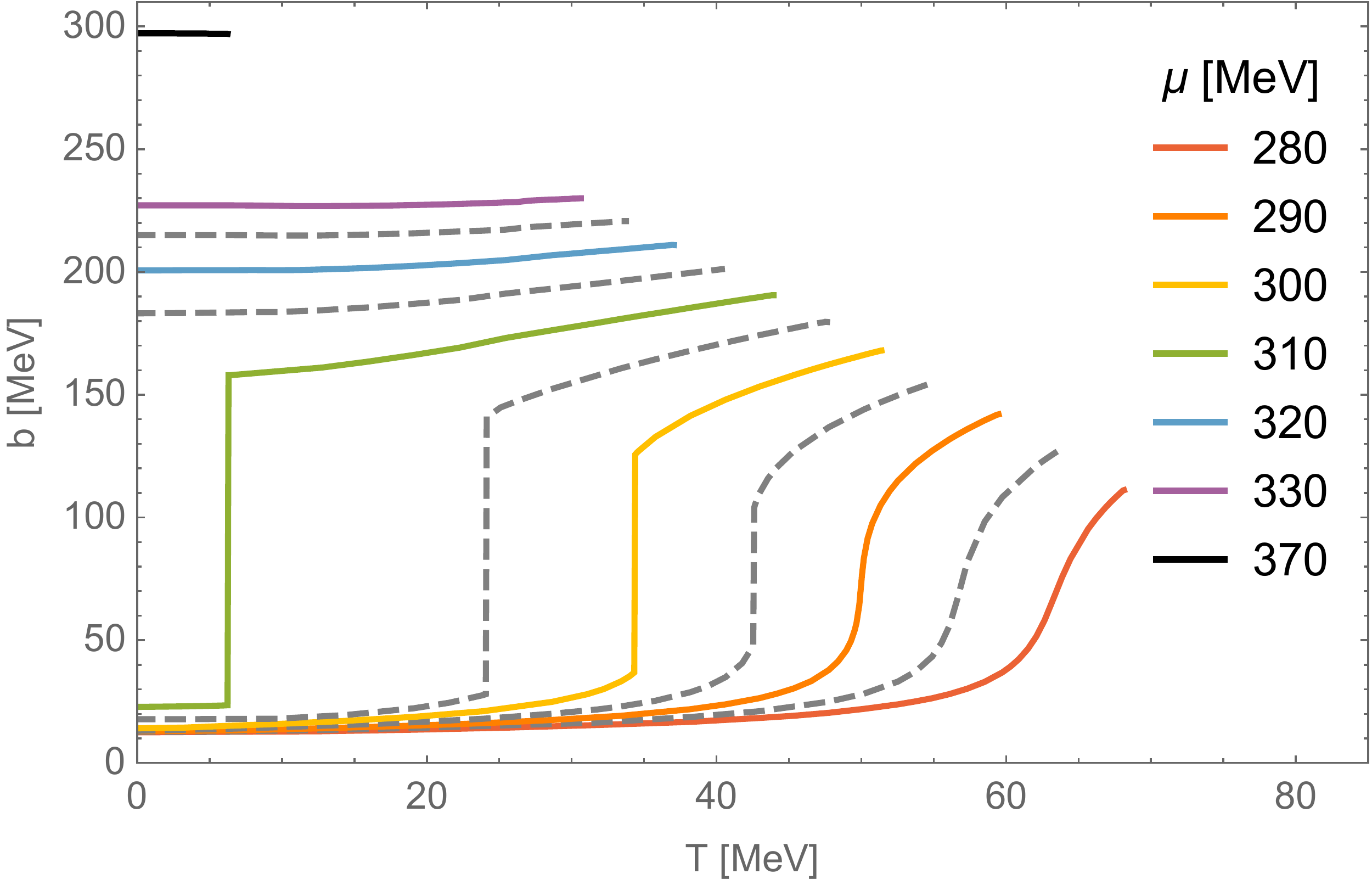}}\hfill
\caption{\textbf{Left:} Order parameters vs $\mu$ at select values of $T$ and fixed $B=1.5\times10^{18}$ G. \textbf{Right:} Order parameters vs $T$ at select values of $\mu$ and fixed $B=1.5\times10^{18}$ G. Gray dashed curves indicate values halfway between colored solid curves. These data are calculated using $30^\text{th}$-order GL expansions, except when $T\leqslant10$ MeV or where $m\gtrsim\mu$, in which case the exact numerical free energy was used.}
\label{fig:OPs at high T}
\end{figure}

Noting that the $20^\text{th}$-order GL expansion accurately describes the condensate at various magnetic fields and temperatures, we now use a $30^\text{th}$-order GL expansion to examine the behavior of the condensate at higher temperatures, where the exact numerical calculation would be computationally burdensome. The left panels of Fig. \ref{fig:OPs at high T} show $m$ and $b$ plotted against $\mu$ at several temperatures, increasing in 5-MeV increments up to $65$ MeV. For $B=1.5\times10^{18}$ G, we already saw in Fig. \ref{fig:OPs at B>0} that the remnant mass is erased when $T=5$ MeV. In Fig. \ref{fig:OPs at high T} we see that increasing $T$ has the effect of decreasing $m$ for all $\mu$, which reduces the range of $\mu$ over which $m>0$. We also see that the first-order transition separating Regions I and II shifts to lower $\mu$ and softens, eventually becoming smooth at $T\approx50$ MeV. On the other hand, the lower panels of Fig. \ref{fig:OPs at high T} show that $b$ increases with $T$ until $m$ vanishes. In Sec. \ref{sec:pairing} we explain this dependence of the inhomogeneity on the temperature in terms of pairing mechanisms.

\begin{figure}
    \centering
\subfloat{\label{sfig:critTa}
\includegraphics[width=.49\textwidth]{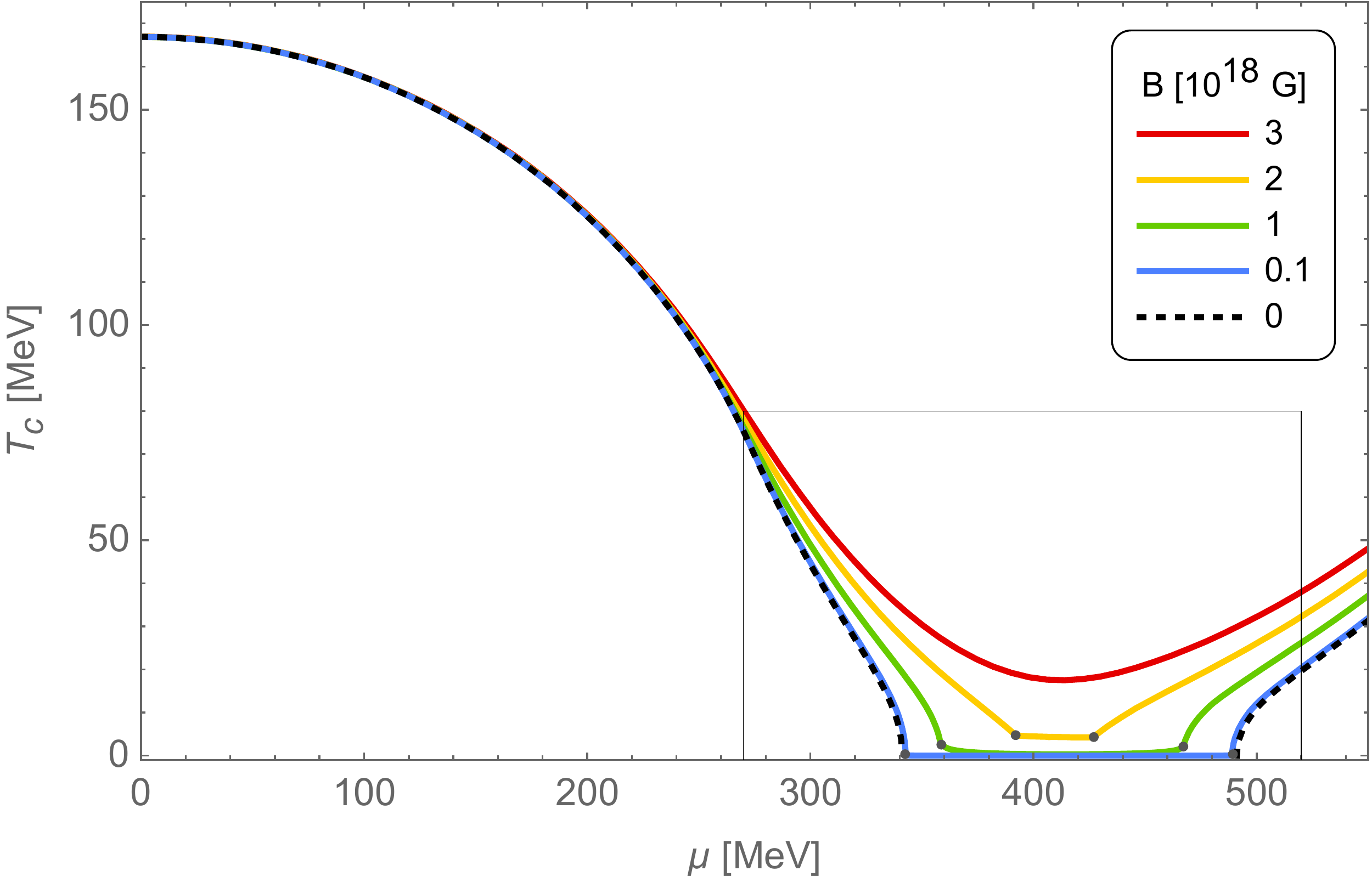}}\hfill
\subfloat{\label{sfig:critTb}
\includegraphics[width=.48\textwidth]{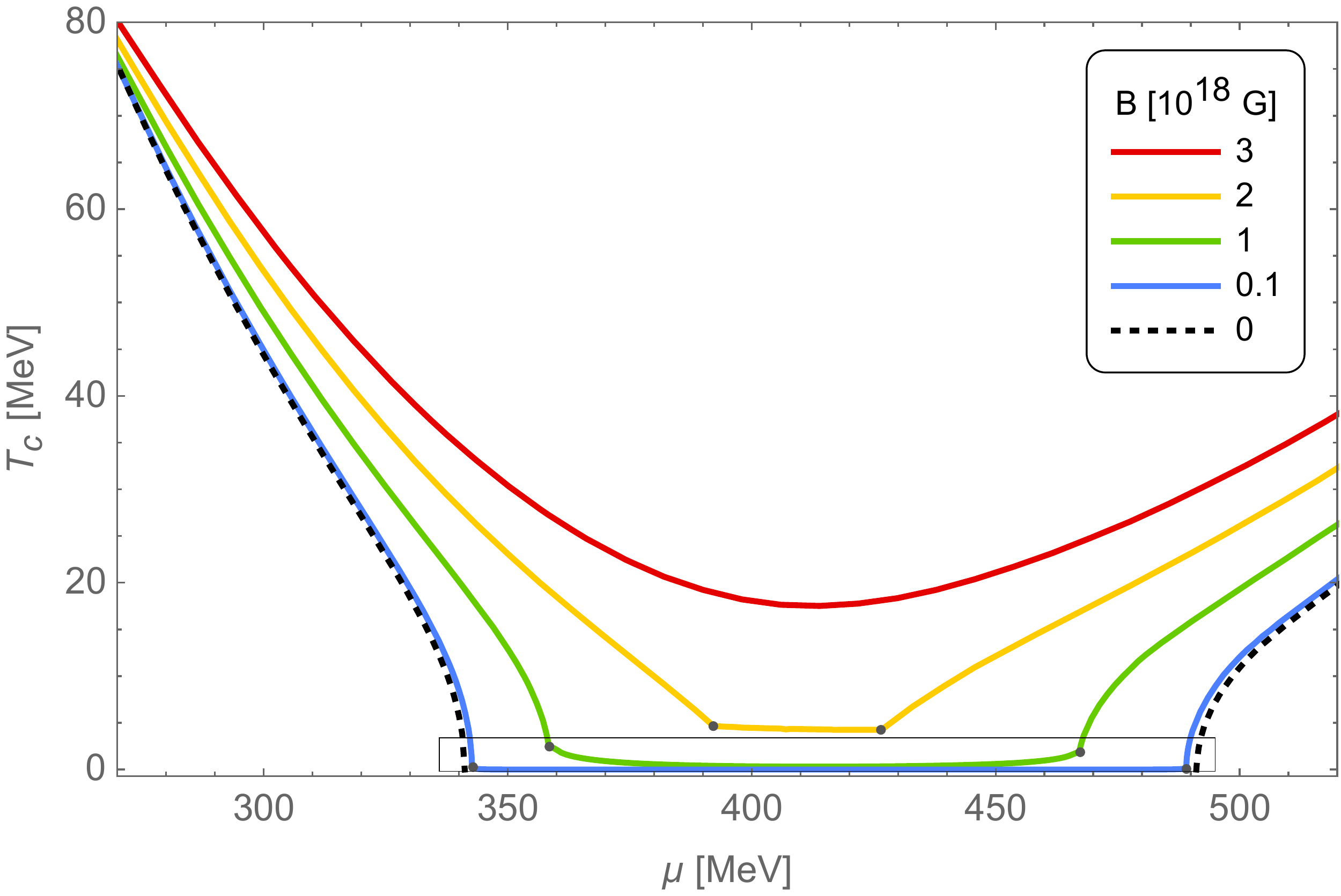}}\hfill
\\
\subfloat{\includegraphics[width=.6\textwidth]{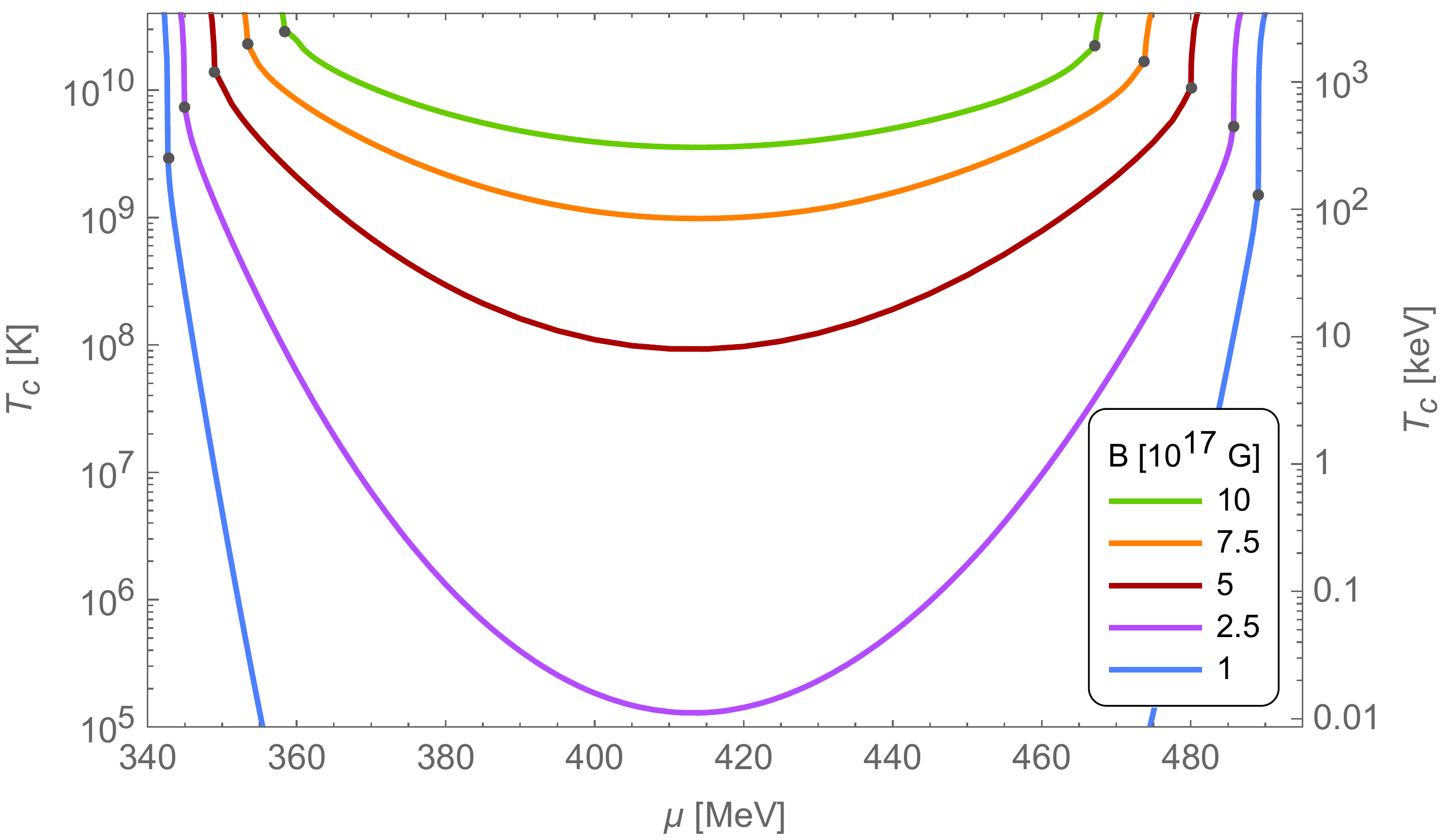}}
    \caption{Critical temperature vs $\mu$ at various magnetic field strengths. Curves indicate the temperature at which $m$ vanishes. Curves are calculated using $30^\text{th}$-order GL expansions to determine where $m>0$ (except in the remnant mass region, where we used a different technique discussed in Appendix \ref{app:remnant T_c}). Gray dots indicate where GL-approximated $T_c$ intersects with the remnant $T_c$. For $B>2\times10^{18}$ G, the remnant region vanishes and is overtaken by the regions of larger $T_c$ on either side.}
    \label{fig:critical T}
\end{figure}

Fig. \ref{fig:OPs at high T} makes apparent that for each $\mu$, there is a critical temperature $T_c$ at which $m$ vanishes. This temperature is plotted in Fig. \ref{fig:critical T} for various magnetic field strengths. Each line in Fig. \ref{fig:critical T} represents the boundary, for a given magnetic field, between the chiral-symmetry-broken inhomogeneous phase (region below) and the chiral-symmetry-restored (region above) phase.  For magnetic fields below $B=2\times10^{18}$ G, we find two regions with large $T_c$, corresponding to the Regions II and IV discussed above, separated by a region with a small remnant $T_c$ required to erase the remnant mass. Apart from the remnant values, the critical temperatures of Fig. \ref{fig:critical T} were calculated using $30^\text{th}$-order GL expansions; the remnant $T_c$ values (which cannot be calculated using the GL expansion, for reasons discussed above) were approximated using a different technique, described in Appendix \ref{app:remnant T_c}. The gray dots in Fig. \ref{fig:critical T} indicate where the remnant $T_c$ intersects the $T_c$ curves of Regions II and IV.

In the bottom panel of Fig. \ref{fig:critical T}, we have zoomed in on the critical temperatures over the remnant region for fields on the order of $10^{17}$ G. It is worth highlighting the potential significance of this result for neutron star applications. Although $T_c$ is small on the scale of MeV in this region, estimates for old NS temperatures fall in the keV range. For these neutron stars, the MDCDW condensate may remain the ground state at much higher densities than predicted. For example, in a star with magnetic fields reaching $5\times10^{17}$ G in the interior, MDCDW would remain the preferred state for all chemical potentials, provided that the temperature is less than $10^8\text{ K}\approx10$ keV; without taking the remnant $T_c$ into account, on the other hand, the condensate would appear to vanish when $350$ MeV$\leqslant\mu\leqslant480$ MeV.

\section{Discussion}
\label{sec:discussion}

\subsection{Pairing mechanisms at different regions}
\label{sec:pairing}

Let us examine the general behavior of the order parameters at zero temperature shown in Figs. \ref{fig:OPs at B=T=0} and \ref{fig:OPs at B>0} from a physical perspective. We begin by describing the $B = 0$ case, and in preparation for our more sophisticated analysis, we first consider the historical picture that involves only homogeneous condensates. In vacuum, chiral symmetry is broken by the pairing of quarks with antiquarks, resulting in the formation of a homogeneous chiral condensate $\langle\bar\psi\psi\rangle$ [\citenum{BubCar}]. A chemical potential does not affect the homogeneous condensate until $\mu$ is large enough to trigger the contribution of the medium (Fermi) term, i.e., when $\mu>E$. At this point, a Fermi surface is formed. Soon after, the system undergoes a first-order phase transition to the chirally symmetric phase. Physically, this indicates that the cost in energy to lift the antiquarks to the Fermi surface to pair with the quarks becomes energetically disfavored. Mathematically, what happens is that the Fermi contribution enters with an opposite sign to the vacuum's. As $\mu$ increases further, this competition between Dirac and Fermi contributions is quickly won by the Fermi part, and the first-order transition occurs.

However, the above description overlooks the fact that once a Fermi surface is formed, a new pairing mechanism becomes possible: Particle-hole pairs can be excited near the Fermi surface with little energy cost. Indeed, at the Fermi surface, quarks and quark-holes that co-move in the same spatial direction exchange only small momenta so that forming a bound state do not cost much energy, in contrast to the quark-antiquark case. The nonvanishing total momentum of these pairs gives rise to spatial inhomogeneity. Because of this new pairing mechanism, the DCDW inhomogeneous condensate is energetically preferable over the chirally restored state at a range of chemical potentials where it was previously thought that chiral symmetry was no longer broken [\citenum{BubCar}]. This scenario describes the parameter behaviors shown in Region II of Fig. \ref{fig:OPs at B=T=0}, where $m$ decreases while $b$ increases, signaling the presence of an inhomogeneous condensate. For chemical potentials just before the onset of the Fermi surface (i.e., just before Region I), the condensate gets contributions only from the Dirac sea (vacuum) term, so in that interval the condensate is homogeneous and constant. When $\mu$ reaches Region I, the Fermi term starts to compete with the Dirac contribution, just as before, forcing the magnitude of the condensate to decrease slightly. The difference is that when the quark-hole pairing is activated, the inhomogeneity arises, and the system reaches Region II instead of the restored phase, where the magnitude $m$ drops to a smaller but nonzero value. In this region, the magnitude $m$ gradually decreases, while the modulation $b$ increases until chiral restoration occur, reaching Region III at a higher chemical potential than in the homogeneous case. 

A magnetic field changes this picture in some remarkable ways. The anomalous term (\ref{anomfreeenergy}), coming from the LLL's asymmetric spectrum, creates a topological contribution to the quark number even in the interval of small $\mu$ where the medium term (\ref{eqn:omega mu HLL}) is still zero; this anomalous contribution favors a nonzero $b$. Because of this unique feature, the density wave phase exists in a magnetic field at arbitrarily small but nonzero chemical potentials \cite{Tatsumi}. This behavior is quite different from the $B=0$ case; as can be seen by comparing the Regions I depicted in Fig. \ref{fig:OPs at B=T=0} and the left panels of Fig. \ref{fig:OPs at B>0}. 

In Region II, the onset of the conventional competition between the Dirac and Fermi contributions forces the dynamical mass $m$ to drop and continue decreasing further, indicating that the quark-antiquark pairing is still operative. However, in contrast to the $B=0$ case, $m$ never reaches zero. Furthermore, other nuances emerge in the density wave condensate due to the difference between the energy modes with spin up and down projections in the direction of the modulation for each HLL. Quarks with the same Landau level $\ell$ but opposite spins have different Fermi surfaces because for each chemical potential, states with the lower energy will be occupied first. Thus, there will be more quarks and hence more holes with the spin projection that corresponds to the lower energy modes at each Landau level.
Given that the chiral condensate comes from quark and holes pairing with opposite spins, the stress created by their different Fermi surfaces adds an element of complexity to the origin of the parameters' behavior with $\mu$ in the inhomogeneous case. Consequently, the particle-hole pairing is not very effective in the range of chemical potentials covering Regions II and III. Such a situation will last until the density becomes large enough to produce a much larger number of quarks and holes of opposite spins available to pair, something that occurs in Region IV, where the two condensate parameters grow with increasing $\mu$, a clear sign that the quark-hole pairing mechanism entirely drives them at the Fermi surface.

An inhomogeneous condensate solution at higher chemical potentials also occurs when $B=0$, as can be seen in Fig. \ref{fig:OPs at B=T=0}. To the best of our knowledge, the return of the condensate at large $\mu$ when $B=0$ has not been included in previous publications; in many cases, plots do not include sufficiently large $\mu$ for the effect to appear [\citenum{Carignano}] [\citenum{NickelPRD}] [\citenum{Nakano}], whereas in Fig. 3 (a) of [\citenum{Frolov}], these high values of $\mu$ are plotted explicitly without any condensate present. The high-$\mu$ condensate appears in Fig. 3 (b) of [\citenum{Frolov}], where $\sqrt{eB}=0.15\Lambda$, and it is claimed to be magnetic in origin. Here, we argue that the effect is driven by the quark-hole pairing mechanism, which becomes significant at large $\mu$ due to the expanded Fermi surfaces. In Sec. \ref{sec:GL analysis}, we supplement this explanation with a more detailed analysis based on the GL coefficients over these regions.

\subsection{Increase of $b$ with $T$}

In the bottom panels of Fig. \ref{fig:OPs at high T}, we see that temperature tends to increase $b$, even though it has the opposite effect on $m$. This observation is consistent with the physical picture described above. At zero temperature, only the lowest energy modes are populated, and the Fermi surface is sharply defined; with the introduction of temperature, some of the energy modes above the Fermi surface become occupied as the set of populated energy modes acquires a statistical spread. Thus, the temperature tends to increase the momenta of the particles near the Fermi surface, which favors the production of particle-hole pairs with greater total momentum. Since this total momentum is the origin of the inhomogeneity, the positive dependence of $b$ on $T$ agrees with the description of the condensate in terms of the particle-hole pairing mechanism.

\subsection{Role of the GL coefficients on the condensate behavior}
\label{sec:GL analysis}

\subsubsection{Role of $\alpha$ coefficients}

In this section, we examine the role of the different types of GL coefficients on the condensate solutions. With that aim, we start by considering the first two lowest orders of the GL expansion ($6^\text{th}$ and $8^\text{th}$), which, while not very accurate, still display the correct qualitative behavior over Regions II--IV at zero and nonzero magnetic field; additionally, they can be easily analyzed quantitatively.

We estimate the order parameters as follows. The free energy is constant along the line $m=0$ in the $m$-$b$ plane, and the first derivative $\partial\Omega/\partial m$ vanishes everywhere because the free energy depends only on $m^2$. If a local minimum exists at some $(m_{min},b_{min}),$ then we expect $\partial^2\Omega/\partial(m^2)$ to be negative at $(0,b_{min})$. Moreover, we can estimate $b_{min}$ as the $b$ that minimizes $\partial^2\Omega/\partial(m^2)$ along the line $m=0$. We can also use the magnitude of $\partial^2\Omega/\partial(m^2)$ as a rough measure of $m$, assuming that stronger negative curvature at $m=0$ results in a minimum that is farther away. This derivative is easy to write using the GL expansion. At the $6^\text{th}$ order expansion, it is
\begin{equation}
\label{eqn:GL 6 derivative}
    \left.
        \frac{\partial\Omega_{GL,6}}{\partial(m^2)}
    \right|_{m=0}
    =\alpha_{2,0}+\beta_{3,1}b+\alpha_{4,2}b^2+\beta_{5,3}b^3+\alpha_{6,4}b^4.
\end{equation}
In the $B=0$ case, the $\beta$ coefficients vanish and (\ref{eqn:GL 6 derivative}) is easily minimized with respect to $b$, giving
\begin{equation}
\label{eqn:GL 6 bmin}
    b_{min}=\sqrt{-\frac{\alpha_{4,2}}{2\alpha_{6,4}}},
\end{equation}
and then we have
\begin{equation}
\label{eqn:GL 6 min derivative}
    \min\left(
    \left.
        \frac{\partial\Omega_{GL,6}^{B=0}}{\partial(m^2)}
    \right|_{m=0}
    \right)
    =\alpha_{2,0}-\frac{\alpha_{4,2}^2}{4\alpha_{6,4}}.
\end{equation}

\begin{figure}[h]
    \centering
\subfloat{
\includegraphics[width=.48\textwidth]{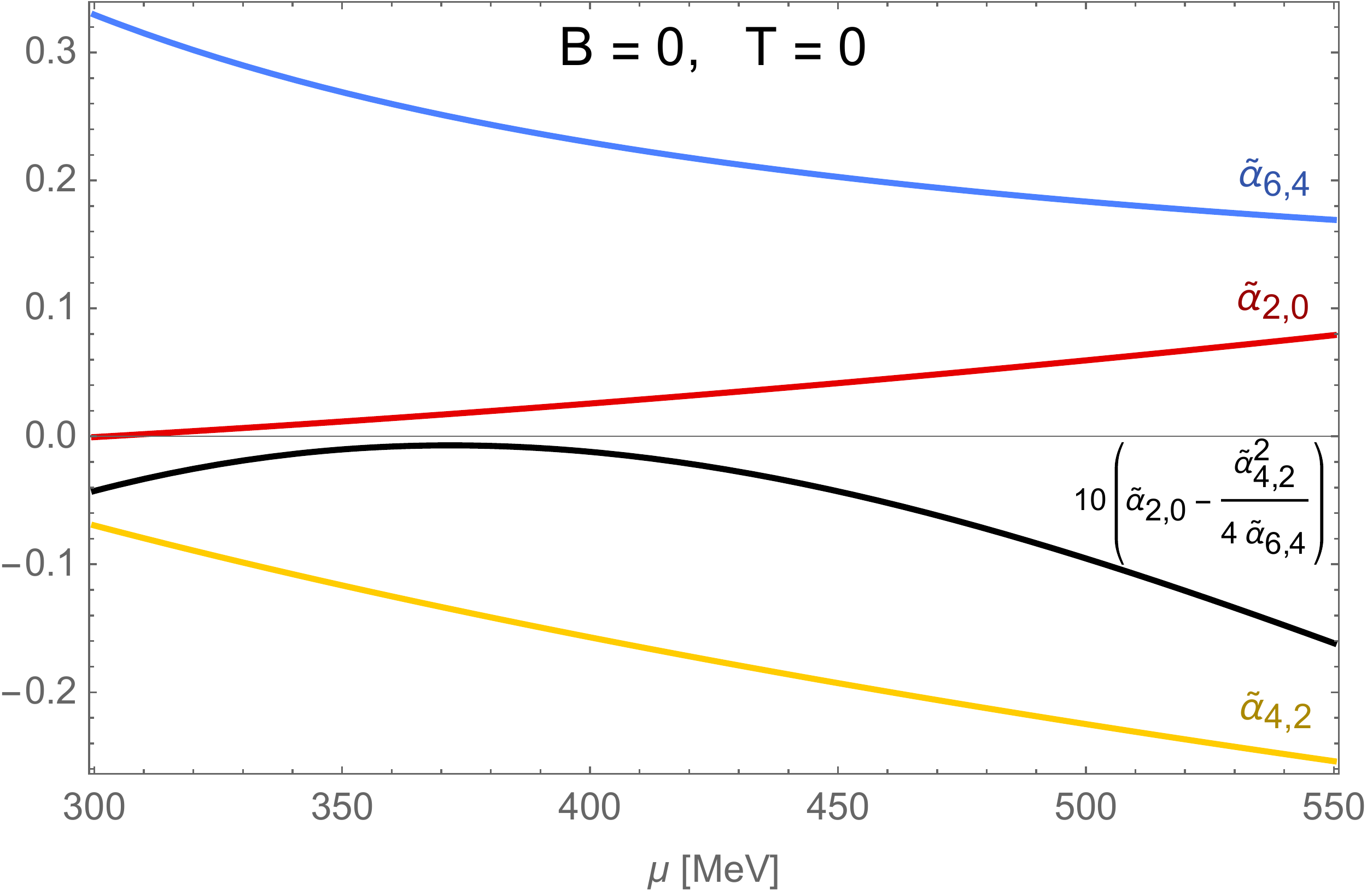}}\hfill
\includegraphics[width=.47\textwidth]{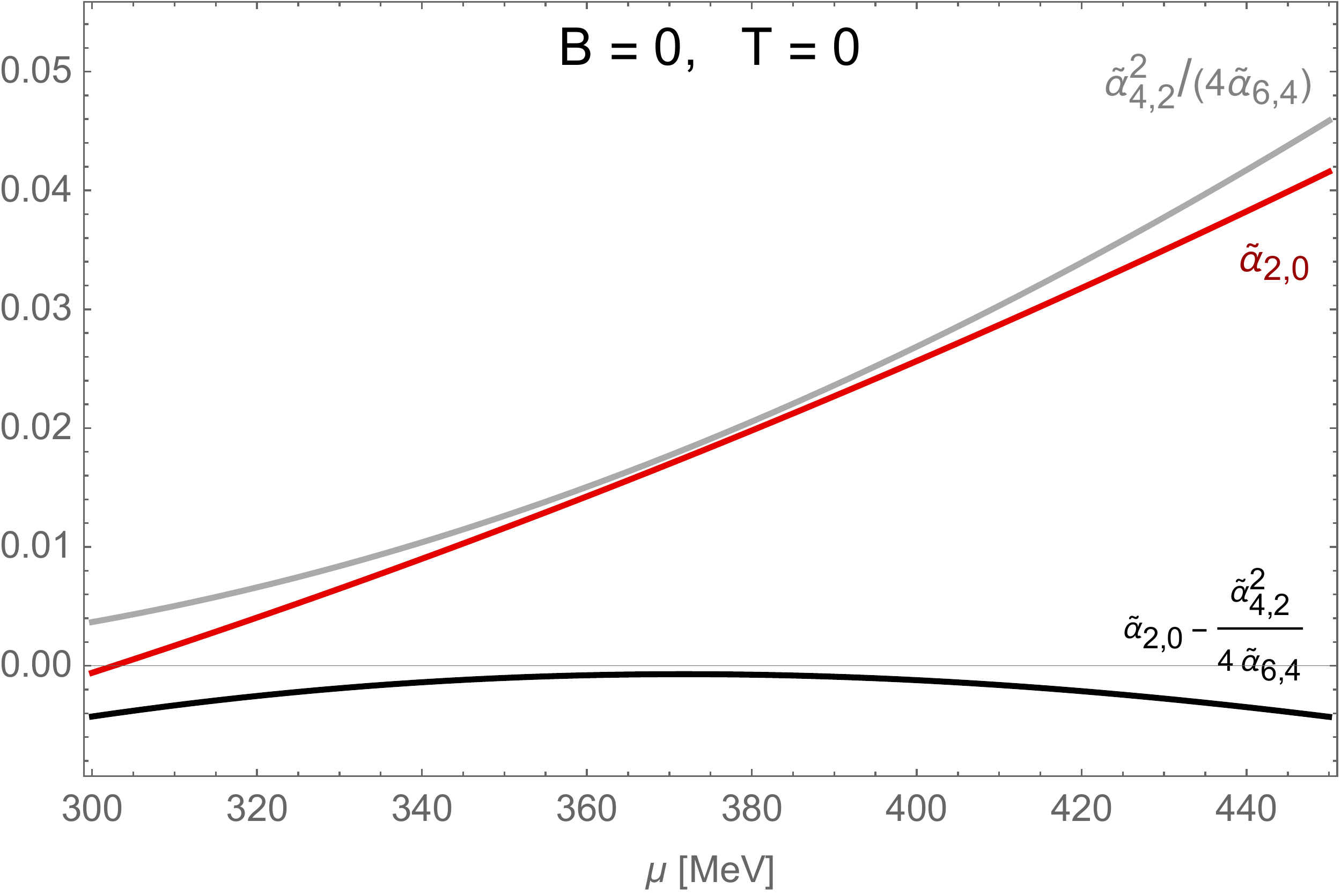}
\\
\subfloat{
\includegraphics[width=.48\textwidth]{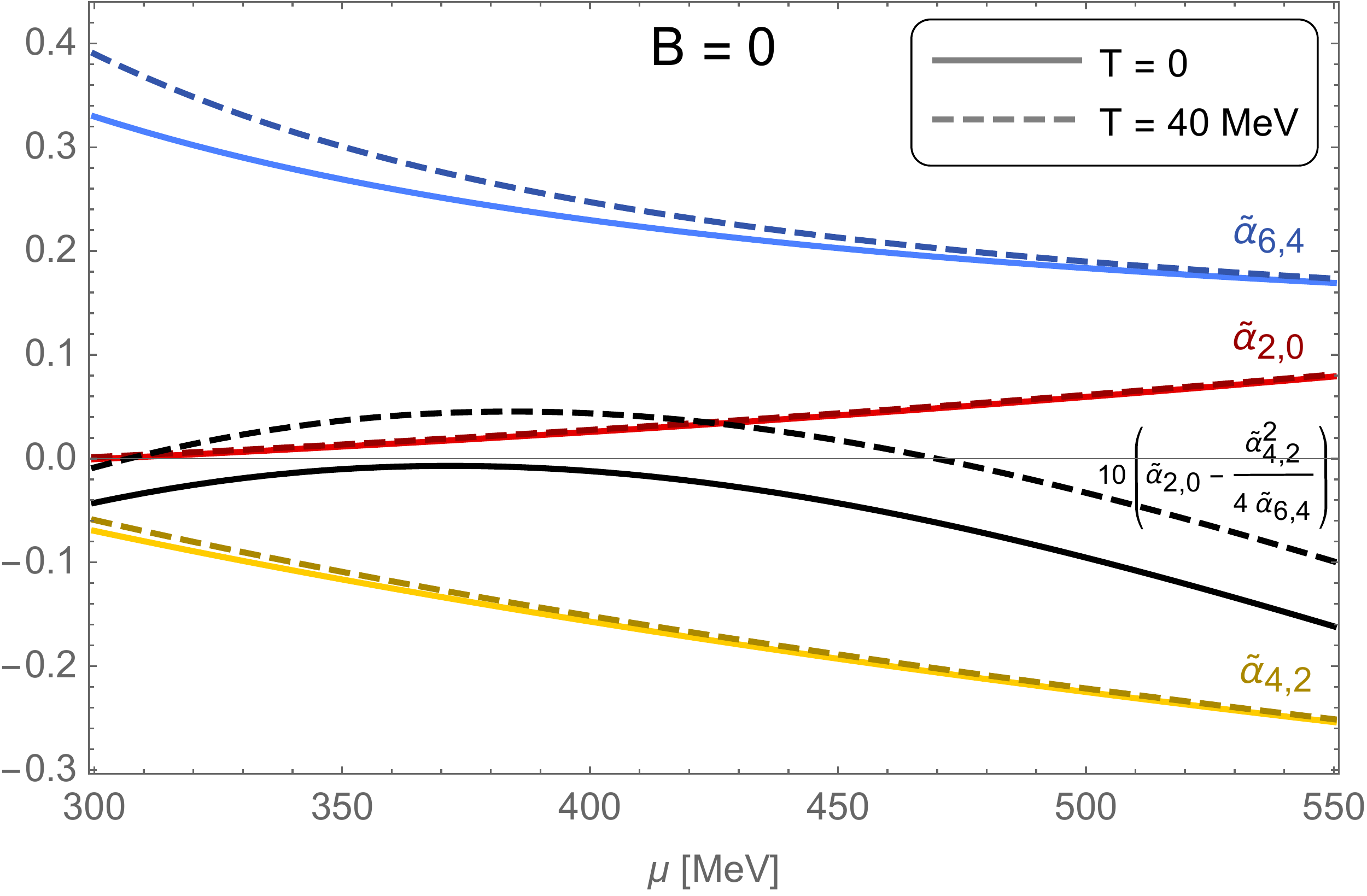}}\hfill
\subfloat{
\includegraphics[width=.49\textwidth]{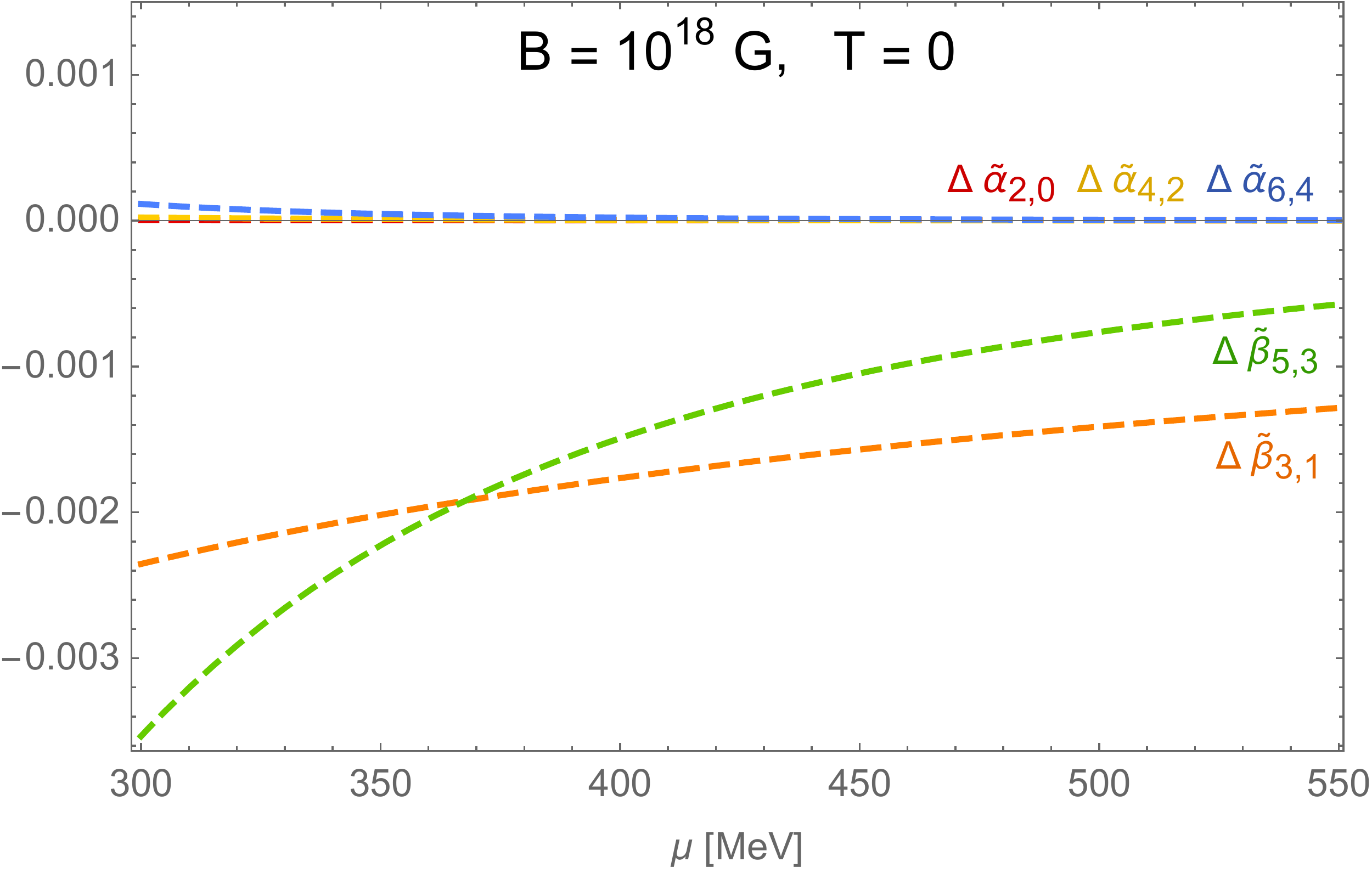}}\hfill
    \caption{Analysis of condensate behavior from GL coefficients. Coefficients are made dimensionless with powers of $\Lambda$, e.g., $\tilde{\alpha}_{2,0}=\alpha_{2,0}/\Lambda^2$. \textbf{Top left:} Coefficients at $B=T=0$. The black curve shows the combination of coefficients that roughly estimates the strength of the condensate, scaled up by a factor of 10 for clarity. More negative values of the black curve correspond to larger values of $m$. \textbf{Top right:} The two competing terms $\alpha_{2,0}$ and $\alpha_{4,2}^2/(4\alpha_{6,4})$, whose difference (black) is related to the condensate strength. For smaller $\mu$, $\alpha_{2,0}$ increases more rapidly, weakening the condensate; at larger $\mu$, the situation reverses and the condensate is strengthened. \textbf{Bottom left:} $\alpha$ coefficients at $T=0$ (solid) and $T=40$ MeV (dashed). Temperature has the effect of ``lifting'' all curves, which corresponds to weakening the condensate. \textbf{Bottom right:} Coefficients at $B=10^{18}$ G, with their values at $B=0$ subtracted off. The $\beta$ coefficients dominate because they scale with $|eB|$, whereas the changes in the $\alpha$ coefficients scale with $|eB|^2$. Since the $\beta$ coefficients are negative, they strengthen the condensate [see Eq. (\ref{eqn:GL 6 derivative}) and preceding discussion].}
    \label{fig:coefs}
\end{figure}

By the above argument, we can treat (\ref{eqn:GL 6 min derivative}) as an estimate of the relative size of $m$, with more negative values indicating larger $m$. The quantity in (\ref{eqn:GL 6 min derivative}), as well as its two components and the individual GL coefficients, are plotted in Fig. \ref{fig:coefs} (top panels). Observe that the gray curve, which corresponds to the quantity in (\ref{eqn:GL 6 min derivative}), is similar to the inversion of the $6^\text{th}$-order curve corresponding to $m$ in Fig. \ref{fig:OPs at B=T=0}, as expected. As $\mu$ increases, $\alpha_{2,0}$ increases, which disfavors the condensate. However, $\alpha_{4,2}$ becomes more negative and $\alpha_{6,4}$ becomes less positive, which both act in favor of the condensate. Eventually, the latter effects dominate, and the condensate magnitude increases, but there is an intermediate region where the condensate is small. Let us recall that the $6^\text{th}$-order GL expansion is not accurate enough to reproduce the actual vanishing of the condensate at intermediate chemical potential when $B=0$.

\begin{figure}[h]
    \centering
\includegraphics[width=.49\textwidth]{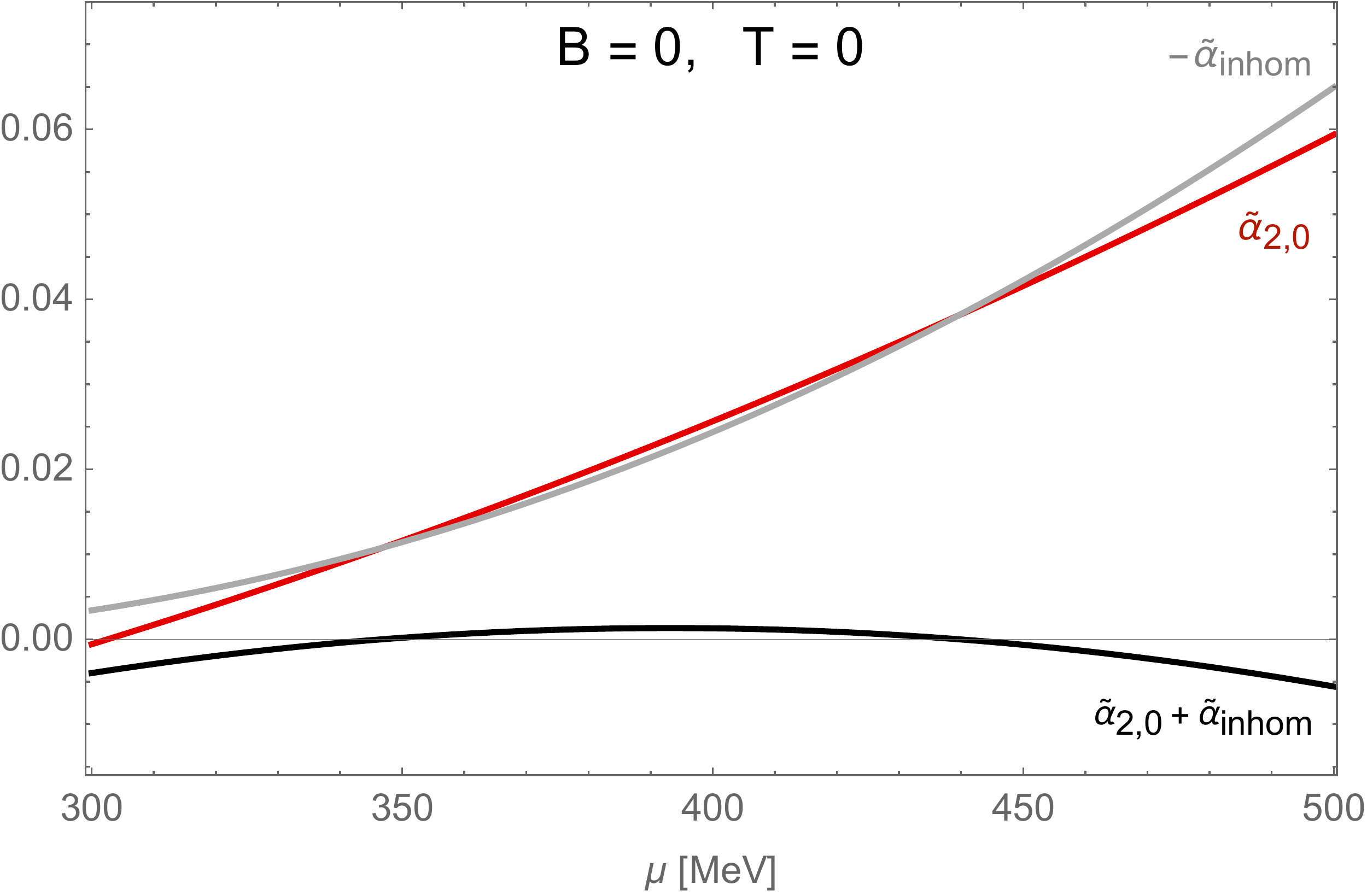}\hfill
    \caption{Analysis of condensate behavior from GL coefficients at $8^\text{th}$ order. As in the top-right panel of Fig. \ref{fig:coefs}, the terms $\alpha_{2,0}$ and $-\alpha_\text{inhom}^{(8)}$ have competing effects on the condensate. Their difference (gray) is related to the condensate strength: more negative values correspond to larger $m$. Unlike in the $6^\text{th}$-order case, here the gray curve is positive over the region $347\text{ MeV}\leqslant\mu\leqslant439$ MeV, implying $m=0$. The full calculation of $m$ using the $8^\text{th}$-order GL expansion predicts $m=0$ over exactly this range (see Fig. \ref{fig:OPs at B=T=0}).}
    \label{fig:8th order}
\end{figure}

The initial analysis at $B=0$ can be repeated at $8^\text{th}$-order, in which $m$ vanishes and reappears. In this case, we have
\begin{align*}
    \left.\frac{\partial\Omega_{GL,8}}{\partial(m^2)}\right|_{m=0}&=\alpha_{2,0}+\alpha_{4,2}b^2+\alpha_{6,4}b^4+\alpha_{8,6}b^6\\\nonumber\\
    &=\alpha_{2,0}+\alpha_\text{inhom}^{(8)},
\end{align*}
where we have defined $\alpha_\text{inhom}^{(8)}\equiv\alpha_{4,2}b^2+\alpha_{6,4}b^4+\alpha_{8,6}b^6$ since this contribution contains all the dependence on the inhomogeneity parameter $b$. It is straightforward to show that  $(b_{min})^2=\frac1{3\alpha_{8,6}}\left(-\alpha_{6,4}+\break\sqrt{\alpha_{6,4}^2-3\alpha_{8,6}\alpha_{4,2}}\right)$, which enables us to recreate the top-right panel of Fig. \ref{fig:coefs} for the $8^\text{th}$-order case, as shown in Fig. \ref{fig:8th order}. Unlike at $6^\text{th}$-order, there is now a region ($347\text{ MeV}\leqslant\mu\leqslant439$ MeV) in which the minimum of $\partial\Omega/\partial(m^2)|_{m=0}$ is positive, which implies the minimum free energy occurs at $m=0.$ Indeed, the $8^\text{th}$-order GL curve (yellow) in Fig. \ref{fig:OPs at B=T=0} vanishes over exactly this range. 

In light of the discussion in the preceding section, we highlight that in this more quantitative analysis, we still find two competing contributions to $m$, only one of which depends directly on the inhomogeneity. The coefficient $\alpha_{2,0},$ which is not associated with any powers of $b$, increasingly disfavors the condensate with increasing $\mu,$ whereas $\alpha_\text{inhom}$, which contains all the coefficients associated with $b$, increasingly favors the condensate. The former effect dominates at smaller $\mu$ (Region II). In contrast, the two effects approximately cancel over intermediate $\mu$ (Region III), and finally, the latter effect dominates at large $\mu$ (Region IV). Therefore, this quantitative analysis from the GL expansion is consistent with our physical interpretation of Dirac and Fermi contributions and pairing mechanisms on each region.

\subsubsection{Magnetic field, role of the $\beta$ coefficients, and remnant mass}
\label{sec:remnant mass}

\begin{figure}[h]
    \centering
\subfloat{
\includegraphics[width=.47\textwidth]{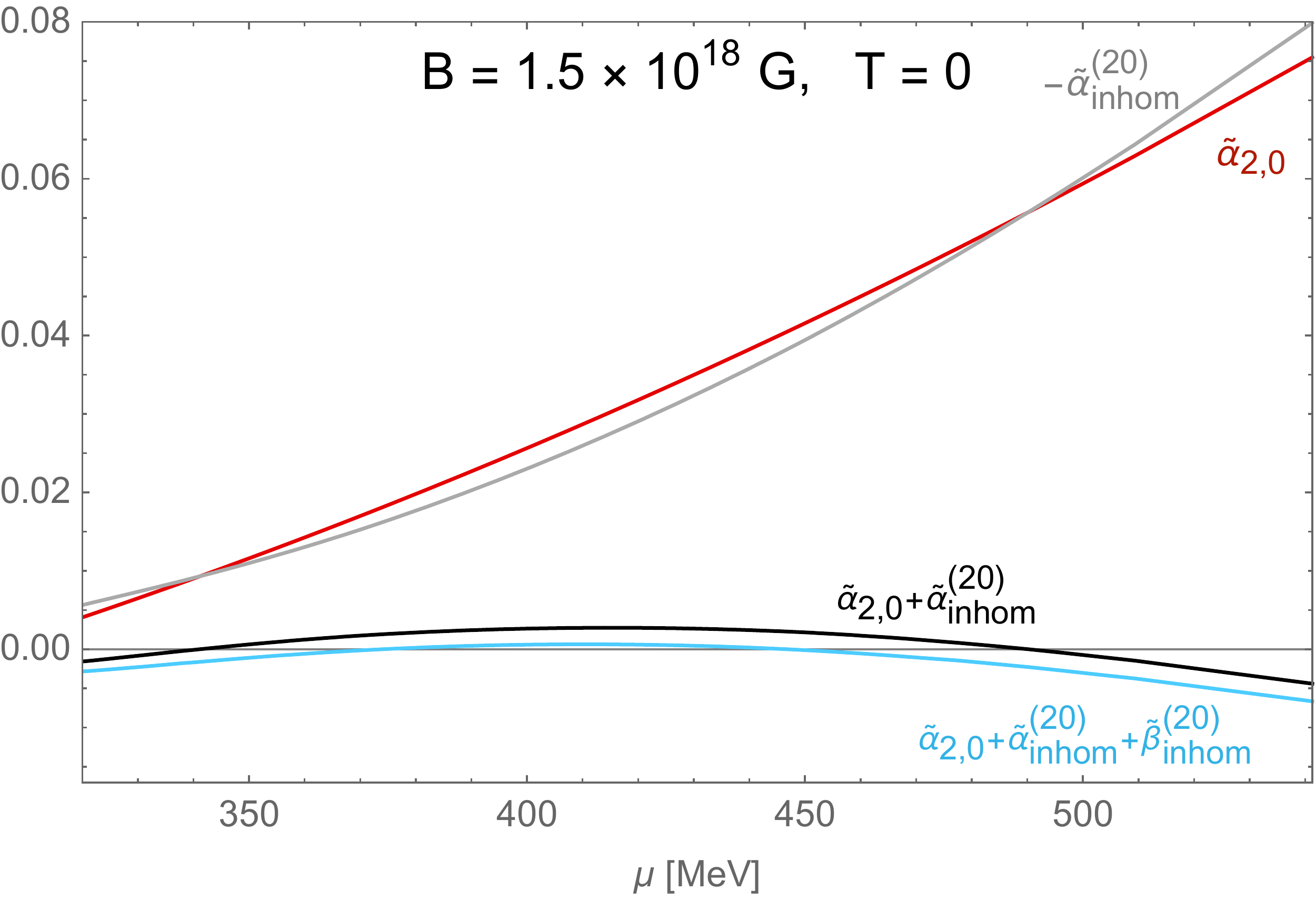}}\hfill
\subfloat{
\includegraphics[width=.48\textwidth]{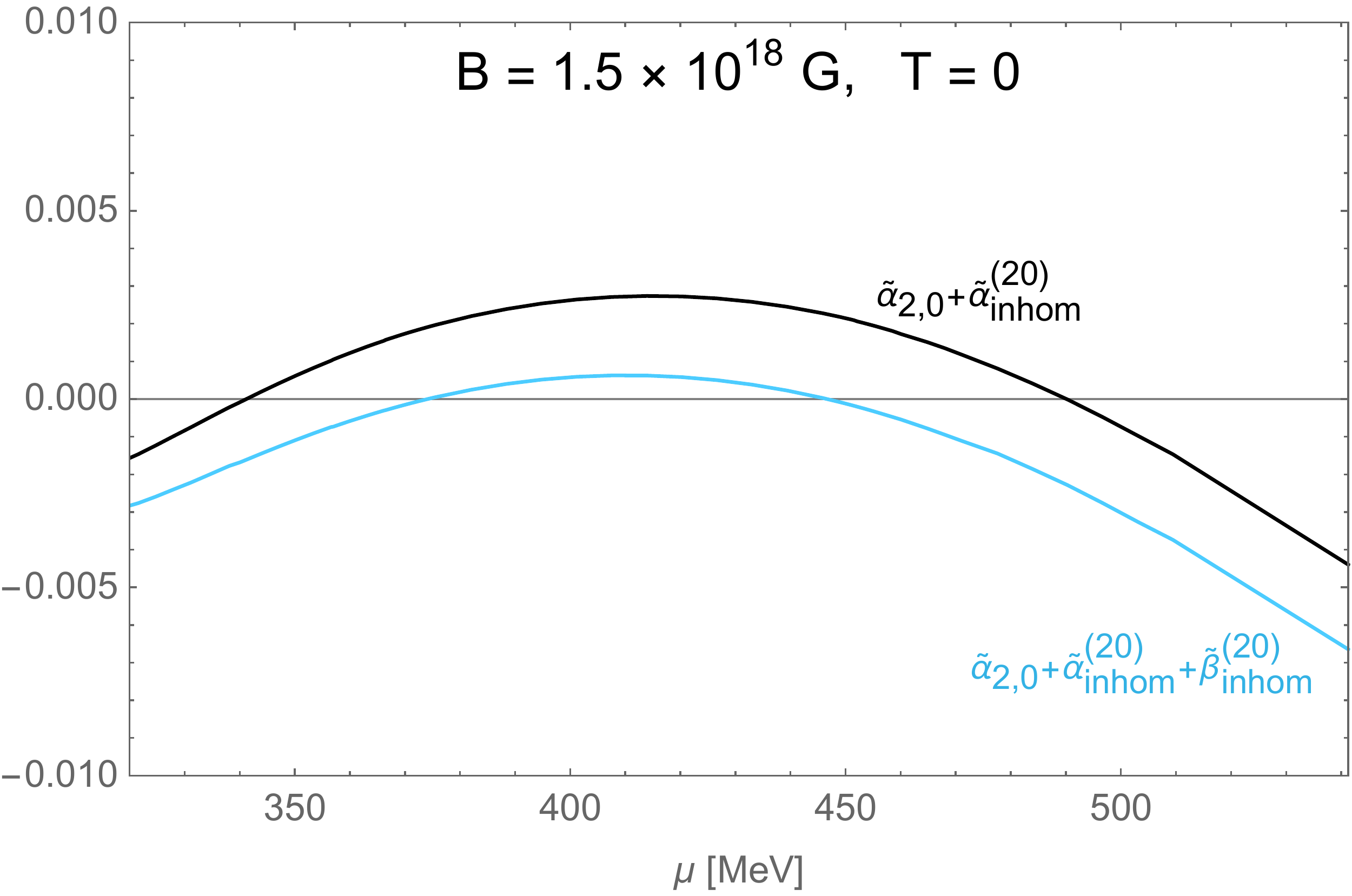}}
\caption{Analysis of condensate behavior from GL coefficients at $20^\text{th}$ order for $B=1.5\times10^{18}$ G. As in Fig. \ref{fig:8th order}, the red, gray, and black curves show various contributions to the derivative of the free energy at $m=0$. Here, the $\beta$ coefficients (not plotted) also contribute to this quantity, with the effect of shifting the derivative (black) to lower values (blue). Unlike Fig. \ref{fig:8th order}, these curves rely on previously calculated values of $b$ that minimize the free energy in this region.}
\label{fig:20th Order}
\end{figure}

Finally, let us examine the effect of the magnetic field on the preceding ``competition" curves and the emergence of the remnant mass. In this case, the $\beta$ coefficients no longer vanish, so we cannot solve analytically for $b_{min}$ using the quadratic formula, as before. Instead, we borrow the numerical results of Fig. \ref{fig:OPs at B>0} (top-left panel), evaluating the derivative at the value of $b$ that minimized the exact numerical free energy (black) at the corresponding chemical potential. Because we cannot solve for $b_{min}$ analytically anyway, we may as well use a very high-order GL expansion for the sake of accuracy. Expanding to $20^\text{th}$ order, we have
\begin{align*}
    \left.\frac{\partial\Omega_{GL,20}}{\partial(m^2)}\right|_{m=0}&=\alpha_{2,0}+\alpha_\text{inhom}^{(20)}+\beta_\text{inhom}^{(20)},
\end{align*}
where $\alpha_\text{inhom}^{(20)}=\alpha_{4,2}b^2+\cdots+\alpha_{20,18}b^{18}$ and $\beta_\text{inhom}^{(20)}=\beta_{3,1}b+\cdots+\beta_{19,17}b^{17}$.  The corresponding curves are plotted in Fig. \ref{fig:20th Order}. Since the $\beta$ coefficients are all negative, they act to decrease the derivative, favoring larger $m$. Note how taking the $\beta$ coefficients into account significantly reduces the range of chemical potential over which $m$ vanishes. In other words, the $\beta$ coefficients favor the emergence of the remnant mass even if they alone are not enough to ensure $m \neq 0$ in the entire remnant mass region found from the exact curve. Since the origin of the $\beta$ coefficients is the asymmetry of the LLL, this indicates that the remnant mass is at least partially a consequence of the topology of the LLL dynamics. 

In addition, and as mentioned before, the remnant mass is also influenced by the behavior of the derivatives in certain regions of the parameter space at each Landau level. To better understand this, let us consider the quantity $\partial\Omega/\partial(m^2)$ along the line $m=0$ in the $m$-$b$ plane. As explained above, if the derivative $\partial\Omega/\partial(m^2)$ is negative anywhere along the line $m=0$, then the minimum free energy must occur at $m>0$. The remnant mass arises because this derivative must be negative at certain values of $b$. In particular, for fixed $B>0$, the contribution to $\partial\Omega/\partial(m^2)$ coming from the $n^\text{th}$ LL (for any allowed $n$) is $-\infty$ at $m=0$, $b=\sqrt{\mu^2-2|e_fB|n}$, as shown rigorously in Appendix \ref{app:remnant mass}. Hence, in Region III, many small local minima exist near the line $m=0$, the deepest of which determines the remnant $m$ and $b$ values. Note that the locations of these singularities along the line $m=0$ depend explicitly on the discretization of Landau levels. Hence, this effect cannot occur in the absence of a magnetic field.

\subsubsection{Opposite effects of magnetic field and temperature on the condensate}

The original analysis at $6^\text{th}$-order can also be extended to explore the effects of temperature and magnetic field on the condensate. The bottom-left panel of Fig. \ref{fig:coefs} compares the $\alpha$ coefficients at $T=0$ and $T=40$ MeV. Each coefficient increases with temperature, which in turn leads to lifting the black curve. This increase reflects the fact that temperature tends to disfavor the condensate.

Alternatively, a strong magnetic field has the effect of making the $\beta$ coefficients more negative (Fig. \ref{fig:coefs}, bottom-right panel). We can see from (\ref{eqn:GL 6 derivative}) that negative $\beta$ coefficients act in favor of the condensate. Interestingly, $B$ has the opposite effect on the $\alpha$ coefficients, increasing them slightly. This difference is negligible compared to that of the $\beta$ coefficients, however, because the $\beta$ coefficients scale with $|eB|$ whereas the $\alpha$ coefficients scale only with $|eB|^2$, which is much smaller (e.g., $10^{18}$ G corresponds to $|eB|/\Lambda^2\approx0.0146$). Since the $\beta$ coefficients only have contributions from the LLL, we can conclude that the LLL drives the main effect of the magnetic field on the condensate, even though the HLLs play a role in the existence of the remnant mass.

Finally, in the $T\to0$ limit of (\ref{eqn:alpha}), one can show that the $\alpha_{n,n-2}$ for $n\geqslant6$ are positive. Hence the higher-order versions of (\ref{eqn:GL 6 derivative}) will approximate a less negative value of $\partial\Omega/\partial(m^2)$, and thus smaller value of $m$. A higher-order expansion will also better approximate the free energy, so we should expect $m$ to decrease by smaller amounts toward some limiting value as the order of the expansion increases. This is exactly what we observe in Figs. \ref{fig:OPs at B=T=0} and \ref{fig:OPs at B>0}.

\section{Conclusion}
\label{sec:conclusion}

This paper investigated the condensate solutions and the phase diagrams at finite temperature and density of the MDCDW phase of dense quark matter using both a generalized GL expansion and exact numerical calculations. To carry out this study, we developed a systematic approach based on mathematical techniques outlined in Appendices \ref{app:coef prefactors} \& \ref{app:alpha}, to derive a set of analytical relations among the GL coefficients that enable their fast computation at arbitrary order. Therefore, this method allows increasing the accuracy of the GL expansion to any desired order with ease, thereby turning the problem of minimizing the free energy (which is challenging to compute numerically when there is a magnetic field) into the minimization of a simple polynomial.

Using both analytical and numerical arguments, we demonstrated that a GL expansion of order 20th or higher accurately approximates the free energy in the region of interest. These results then provide a pretty reliable tool to study the effects of the magnetic field, temperature, and density on the condensate. Using them, we showed that magnetic fields of magnitude $10^{17}$ G and higher noticeably increase the condensate magnitude and expand the region of phase space in the $\mu$-$T$ plane where the condensate is preferred. As expected, high temperatures produce the opposite effect, so they tend to decrease the magnitude of the inhomogeneous condensate and reduce the region of chemical potentials where it is preferred.

 A significant outcome of our study has been uncovering the resilience of the inhomogeneous chiral condensate in regions of densities where previous studies had found it to vanish. This resilience manifests itself in two important senses. The first sense is the phenomenon we have called the ``remnant mass," which occurs only in the presence of a magnetic field. The remnant mass manifests as an inhomogeneous chiral condensate with small but nonvanishing magnitude $m$ and significant modulation $b$ in the intermediate $\mu$ region. The second sense is the reappearance of sizable $m$ and $b$ at very high chemical potentials and their increase with $\mu$. This one happens with and without a magnetic field. These effects may have significant consequences for astrophysical applications because the broadened range of conditions in which the condensate is supported may overlap with those realized in compact stars. 

The developed computational tools enabled unveiling the properties mentioned above with straightforward calculations. On the other hand, the physical mechanisms underlying the behavior of the condensate with and without a magnetic field are rather complex. For example, some features remain qualitatively unaffected by a magnetic field, such as the resurgence of $m$ at large $\mu$. In contrast, other features, such as the inhomogeneity at small $\mu$ and the remnant mass at intermediate $\mu$, only appear when a magnetic field is present. Remarkably, these last two features are connected to the nontrivial topology of the LLL in the MDCDW phase. In addition, the discretization of the HLL also plays a role in the latter. Finally, much of this behavior can be understood in terms of the quark-antiquark and quark-hole pairing that drives the condensate formation. These pairing mechanisms have competing effects, the consequences of which are far from obvious.

A meaningful extension of this work will be quantitatively analyzing the condensate's stability against thermal fluctuations at temperatures relevant to neutron star applications. The present study has found regions of temperature where the inhomogeneous condensate is preferred at various magnetic fields and chemical potentials that are well within the range of neutron star temperatures. Still, even at temperatures where MDCDW is energetically favored in mean-field theory, it is unclear if the thermal fluctuations could erase the long-range correlations it creates. In the absence of a magnetic field, we know that single-modulated condensates are unstable against such fluctuations at arbitrarily small temperatures, a phenomenon known as the Landau-Peierls instability. It was shown in [\citenum{AbsenceLP}], however, that this phenomenon does not occur when a magnetic field is present. Thus, a natural task is determining the threshold temperature at which the fluctuations would erase the condensate under various conditions. The current work shows that if such a threshold temperature is sufficiently high, then the MDCDW condensate may be the preferred state over an extensive range of conditions, enhancing the likelihood of this phase being realized in the extreme environments of compact stars.

\section*{ACKNOWLEDGMENTS}

We thank Efrain J. Ferrer, Sarang Gopalakrishnan, Vadim Oganesyan, Rob Pisarski, and Israel Portillo for valuable discussions and comments. This work was supported in part by NSF grant PHY-2013222. W.G. is grateful to UTRGV for monetary support during a portion of this work.

\appendix

\section{Relationship between coefficients of equal order}
\label{app:coef prefactors}

Let us denote a general term in the GL expansion as $c_{n,n_b}m^{n_m}b^{n_b}$, where $n=n_m+n_b$. We assume $n\geqslant4$, as otherwise there is only one coefficient of order $n$, in which case the formula we are about to derive is not relevant. The symmetries of the system only allow for even powers of $m$, so we can assume $n_m$ is even and consider derivatives with respect to $m^2$ rather than $m$. We immediately have
\begin{equation}
	\label{coef defn}
	c_{n,n_b}=\lim_{m,b\to 0}\frac{1}{(n_m/2)!\,n_b!}
	\left(\frac{\partial}{\partial(m^2)}\right)^{n_m/2}
	\left(\frac{\partial}{\partial b}\right)^{n_b}\Omega.
\end{equation}
We now make use of the crucial observation that the free energy can be expressed as
\begin{equation}
	\label{omega integral form}
	\Omega=\int_{-\infty}^{+\infty}dk\,[f(E_+^0)+f(E_-^0)],
\end{equation}
where $E^0_\pm=\pm\sqrt{m^2+k^2}+b$ and $f$ is some sufficiently well-behaved function. Specifically, we assume that $f$ is real-analytic and vanishes rapidly at infinity. It is possible to verify these properties by manipulating the free energy given in (\ref{eqn:free energy})--(\ref{eqn:omega mu LLL}) to find an explicit expression for $f$. Still, the precise function is not relevant here. Note that although $E^0$ is the energy of the lowest Landau level, the integrand of (\ref{omega integral form}) also includes the higher Landau levels, since the energies of the HLL are also functions of $E^0$. We neglect the $m^2/(4G)$ term because it only affects the coefficient $\alpha_{2,0}$, whereas we are considering only $n\geqslant4.$

Let us use the condensed derivative notation $\partial_x=\partial/\partial x$ and suppress the limits of integration. Inserting (\ref{omega integral form}) in (\ref{coef defn}) we have
\begingroup
\allowdisplaybreaks
\begin{align}
	c_{n,n_b}&=\lim_{m,b\to 0}\frac{1}{(n_m/2)!\,n_b!}
	\partial_{m^2}^{n_m/2}
	\partial_b^{n_b}
	\int dk\,[f(E_+^0)+f(E_-^0)]
	\nonumber\\\nonumber\\
	&=\lim_{m,b\to 0}\frac{1}{(n_m/2)!\,n_b!}
	\partial_b^{n_b}
	\int dk\,
	\partial_{m^2}^{n_m/2}
	[f(E_+^0)+f(E_-^0)]
	\nonumber\\\nonumber\\
	&=\lim_{m,b\to 0}\frac{1}{(n_m/2)!\,n_b!}
	\partial_b^{n_b}
	\int dk
	\left(\frac{1}{2k}\partial_k\right)^{n_m/2}
	[f(E_+^0)+f(E_-^0)]
\end{align}
\endgroup
where in the last step we use the fact that $\partial_{m^2}\sqrt{m^2+k^2}=\frac1{2k}\partial_k\sqrt{m^2+k^2}.$ We note that many of the commutations of limits, derivatives, and integrals throughout this calculation require careful dominated and uniform convergence proofs to be fully justified. These proofs are possible, aided by the tameness properties of $f$ and symmetries of the integrand. 

Now taking the limit $m\to0$ under the integral sign, we have
\begingroup
\allowdisplaybreaks
\begin{align}
	c_{n,n_b}&=\lim_{b\to 0}\frac{1}{(n_m/2)!\,n_b!\,2^{n_m/2}}
	\partial_b^{n_b}
	\int dk
	\left(\frac{1}{k}\partial_k\right)^{n_m/2}
	[f(k+b)+f(-k+b)]
	\nonumber\\\nonumber\\
	&=\lim_{b\to 0}\frac{1}{(n_m/2)!\,n_b!\,2^{n_m/2}}
	\int dk
	\left(\frac{1}{k}\partial_k\right)^{n_m/2}
	\partial_b^{n_b}
	[f(k+b)+f(-k+b)]
	\nonumber\\\nonumber\\
	&=\lim_{b\to 0}\frac{1}{(n_m/2)!\,n_b!\,2^{n_m/2}}
	\int dk
	\left(\frac{1}{k}\partial_k\right)^{n_m/2}
	\partial_k^{n_b}
	[f(k+b)+(-1)^{n_b}f(-k+b)]
	\nonumber\\\nonumber\\
	\label{pre IBP}
	&=\frac{1}{(n_m/2)!\,n_b!\,2^{n_m/2}}
	\int dk
	\left(\frac{1}{k}\partial_k\right)^{n_m/2}
	\partial_k^{n_b}
	[f(k)+(-1)^{n_b}f(-k)]
\end{align}
\endgroup

Note that there is no pole at $k=0$ because the term in square brackets is even [odd] if $n_b$ is even [odd], so after acting with $\partial_k^{n_b}$ the result is even, and then acting with $(1/k)\partial_k$ any number of times gives a bounded even function of $k$. Thus we are justified in using integration by parts (IBP). Using IBP $n_m/2$ times gives
\begingroup
\allowdisplaybreaks
\begin{align}
	c_{n,n_b}&=\frac{1}{(n_m/2)!\,n_b!\,2^{n_m/2}}
	\int dk\,(n_m-1)!!
	\frac{1}{k^{n_m}}
	\partial_k^{n_b}
	[f(k)+(-1)^{n_b}f(-k)]
	\nonumber\\\nonumber\\
	&=\frac{(n_m-1)!!}{(n_m/2)!\,n_b!\,2^{n_m/2}}
	\int dk
	\left[\frac{-1}{(n_m-1)!}\partial_k^{n_m-1}\frac{1}{k}\right]
	\partial_k^{n_b}
	[f(k)+(-1)^{n_b}f(-k)]
	\nonumber\\\nonumber\\
	\label{post IBP}
	&=\frac{1}{(n_m/2)!\,n_b!\,2^{n_m/2}(n_m-2)!!}
	\int dk\,
	\frac{1}{k}\,
	\partial_k^{n_m+n_b-1}
	[f(k)+(-1)^{n_b}f(-k)],
\end{align}
\endgroup
where we used IBP again $n_m-1$ times in the last step. 

Now let us consider the coefficient $c_{n,n-2}$. If we repeat the previous calculation for this coefficient, then we will arrive at (\ref{pre IBP}) with $n_m\to2$ and $n_b\to n-2,$ giving
\begingroup
\allowdisplaybreaks
\begin{align}
	\label{no IBP}
	c_{n,n-2}&=\frac{1}{(n-2)!\,2}
	\int dk\,
	\frac{1}{k}
	\partial_k^{n-1}
	[f(k)+(-1)^{n-2}f(-k)].
\end{align}
\endgroup
The integral in (\ref{no IBP}) is identical to that of (\ref{post IBP}) because $n=n_m+n_b$ and also the exponents of $-1$ differ by $n_m-2,$ which is even. Thus, dividing (\ref{no IBP}) by (\ref{post IBP}) and then rearranging gives (\ref{eqn:coef prefactors}).

\section{Calculation of $\bm{\alpha}$ coefficients}
\label{app:alpha}

Let us sketch the derivation of (\ref{eqn:alpha}). Focusing first on the vacuum term and letting $a=2|e_f B|$, from (\ref{eqn:free energy}) we can calculate
\begin{align}
    \lim_{m\to0}\frac{\partial\Omega_{vac}^{f,HLL}}{\partial(m^2)}=-\frac{N_c}{(2\pi)^2}\frac a2\sum_{\ell=1}^\infty\operatorname{PV}
    \int_{-\infty}^{+\infty}dk
    \frac 1{k-b}\frac k{\sqrt{k^2+a\ell}}
    \operatorname{erfc}
    \left(
        \frac{\sqrt{k^2+a\ell}}\Lambda
    \right),
\end{align}
where $\operatorname{PV}\int_{-\infty}^{+\infty}dk=\lim_{\epsilon\to0}(\int_{-\infty}^{b-\epsilon}+\int_{b+\epsilon}^{+\infty})dk$. If we also take $n_b$ derivatives with respect to $b$, we have
\begin{align}
    \lim_{m\to0}
    \frac{\partial^{n_b}}{\partial b^{n_b}}
    \frac{\partial\Omega_{vac}^{f,HLL}}{\partial(m^2)}=-\frac{N_c}{(2\pi)^2}\frac a2
    \sum_{\ell=1}^\infty\operatorname{PV}
    \int_{-\infty}^{+\infty}dk
    &\left[
        \frac{\partial^{n_b}}{\partial b^{n_b}}
        \left(
        \frac 1{k-b}
        \right)
    \right]
    \frac k{\sqrt{k^2+a\ell}}
    \operatorname{erfc}
    \left(
        \frac{\sqrt{k^2+a\ell}}\Lambda
    \right).
\end{align}
Note that we can replace $\partial/\partial b$ with $-\partial/\partial k$. Then taking the limit $b\to0$ gives
\begin{align}
    \lim_{m,b\to0}
    \frac{\partial^{n_b}}{\partial b^{n_b}}
    \frac{\partial\Omega_{vac}^{f,HLL}}{\partial(m^2)}=-\frac{N_c}{(2\pi)^2}\frac a2
    \sum_{\ell=1}^\infty\operatorname{PV}
    \int_{-\infty}^{+\infty}dk
    &\left[(-1)^{n_b}
        \frac{\partial^{n_b}}{\partial k^{n_b}}
        \frac1k
    \right]
    \frac k{\sqrt{k^2+a\ell}}
    \operatorname{erfc}
    \left(
        \frac{\sqrt{k^2+a\ell}}\Lambda
    \right).
\end{align}
If $n_b$ is odd, the integrand is odd. Hence the total expression vanishes as expected. So henceforth, we assume $n_b$ is even. We can integrate by parts $n_b$ times; each time, the boundary terms at $k=\pm\epsilon$ cancel because they are evaluated at an odd function of $k$. Thus we have
\begin{align}
    \lim_{m,b\to0}
    \frac{\partial^{n_b}}{\partial b^{n_b}}
    \frac{\partial\Omega_{vac}^{f,HLL}}{\partial(m^2)}=-\frac{N_c}{(2\pi)^2}\frac a2
    \sum_{\ell=1}^\infty
    \int_{-\infty}^{+\infty}dk &\frac1k
    \frac{\partial^{n_b}}{\partial k^{n_b}}
    \left[
        \frac k{\sqrt{k^2+a\ell}}
    \operatorname{erfc}
    \left(
        \frac{\sqrt{k^2+a\ell}}\Lambda
    \right)
    \right],
\end{align}
Where we have removed the PV because $1/k$ now multiplies an odd function of $k$, so there is no pole at $k=0$.

Now let us assume $T>0$. Then the integrands for the HLL contributions of $\Omega_\mu^f$ and $\Omega_T^f$ in (\ref{eqn:omega mu HLL})--(\ref{eqn:omega T}) can be combined into a single expression, given explicitly in (\ref{eqn:omega mu+T}), to which we can apply a procedure similar to that used above for $\Omega_{vac}^{f,HLL}$. Combining both of these expressions gives
\begin{align}
\label{HLL derivatives}
    \lim_{m,b\to0}
    \frac{\partial^{n_b}}{\partial b^{n_b}}
    \frac{\partial\Omega^{f,HLL}}{\partial(m^2)}=-\frac{N_c}{(2\pi)^2}
    \int_0^\infty dk\frac1k
    \frac{\partial^{n_b}}{\partial k^{n_b}}
    \left[
        \sum_{\ell=1}^\infty
        \frac{a k}{\sqrt{k^2+a\ell}}
        f\Big(\sqrt{k^2+a\ell}\Big)
    \right],
\end{align}
where $\Omega^{f,HLL}=\Omega_{vac}^{f,HLL}+\Omega_\mu^{f,HLL}+\Omega_T^{f,HLL}$ and
\begin{align}
    f(x)\equiv\operatorname{erfc}(x/\Lambda)
    -\frac1{1+e^{\beta(x-\mu)}}
    -\frac1{1+e^{\beta(x+\mu)}}.
\end{align}
Applying the Euler-Maclaurin formula to the term in square brackets, we can write
\begin{align}
    \lim_{m,b\to0}
    \frac{\partial^{n_b}}{\partial b^{n_b}}
    \frac{\partial\Omega^{f,HLL}}{\partial(m^2)}
    \sim\sum_{j=0,1,2,4,6,\ldots}C_{n_b}^{(j)}
    ,
\end{align}
and then proceed to find explicit expressions for the $C_{n_b}^{(j)}$. Defining $f^{(-1)}(x)\equiv\int_\infty^x f(t)dt$ and using the fact that $f$ is an odd function, it is then straightforward to show, applying the Euler-Maclaurin formula,
\begin{align}
\label{j=0 terms}
    C_0^{(0)}&=2\frac{N_c}{(2\pi)^2}
    \int_0^\infty dk f^{(-1)}(k)dk
    \nonumber\\\nonumber\\
    C_{n_b>0}^{(0)}&=2n_b\frac{N_c}{(2\pi)^2}
    \int_0^\infty dk\frac1k f^{(n_b-2)}(k)dk.
\end{align}
To find the expressions for $C_{n_b}^{(j>0)}$, first rewrite the summand in (\ref{HLL derivatives}) as $a(\partial/\partial k)f^{(-1)}(\sqrt{k^2+a\ell})$ and observe that each term $C_{n_b}^{(j>0)}$ involves $j-1$ derivatives with respect to $\ell.$ Following the strategy used in Appendix \ref{app:coef prefactors}, we can use $\partial/\partial\ell=(1/2k)\partial/\partial k$ and then integrate by parts repeatedly to arrive at
\begin{align}
\label{j>0 terms}
    C_{n_b}^{(j>0)}=\left(\frac a2\right)^n
    \frac{2n_b!!}{(n_b+2j-2)!!}
    \frac{B_j^+}{j!}\frac{N_c}{(2\pi)^2}
    \int_0^\infty dk\frac1k f^{(n_b+2j-2)}(k).
\end{align}
The integrals in (\ref{j=0 terms}) and (\ref{j>0 terms}) can all be solved analytically, and it can be shown by direct calculation that the $j=1$ term exactly cancels the total LLL contribution for each $n_b$. Finally, inserting the pre-factor of $1/(n_b!)$, setting $N_c=3$, and summing in flavor gives (\ref{eqn:alpha}).

\section{HLL influence on the remnant mass origin}
\label{app:remnant mass}

In this appendix, we present an analytical explanation of the small but nonzero value of $m$ that persists at intermediate values of $\mu$ for large $B$ and small $T$, which we have called the remnant mass. Fix $eB>0,T=0,\mu>\sqrt{2eB/3},$ and any integer $n$ such that $0<n<\mu^2/(2eB/3)$. Defining $\mu_n\equiv\sqrt{\mu^2-2(eB/3)n}$, we will show
\begin{equation}
\label{curvature limit full}
    \lim_{\epsilon\to0}
    \left(\left.
    \frac{\partial\Omega}
    {\partial(m^2)}
    \right|_{m=0,
    b=\mu_n+\epsilon}
    \right)
    =-\infty.
\end{equation}
In particular, the curvature must be negative along some interval on the line $m=0$ near $b=\mu_n$. Since $\partial\Omega/\partial m=0$ along the line $m=0$, negative curvature implies that $\Omega$ must be decreasing in the direction of increasing $m$ over some region. Thus the (constant) value of $\Omega$ along $m=0$ cannot be the global minimum. 

We will show in particular that the infinite result of the limit in (\ref{curvature limit full}) arises from the $n^\text{th}$ Landau level of the $\Omega_\mu$ contribution, that is,
\begin{equation}
\label{curvature limit component}
    \lim_{\epsilon\to0}
    \left(\left.
    \frac{\partial\Omega_\mu^{f=1,\ell=n}}
    {\partial(m^2)}
    \right|_{m=0,
    b=\mu_n+\epsilon}
    \right)
    =-\infty.
\end{equation}
As we will see, if $n$ is even, then the $f=2$ component of the $(n/2)^\text{th}$ LL also results in $-\infty$ in this limit. It can be shown by direct calculation that the other LLs and components of $\Omega$ contribute only finite amounts to this limit; hence (\ref{curvature limit component}) is equivalent to (\ref{curvature limit full}).

From (\ref{eqn:omega mu HLL}) we can write the zero-temperature fermion contribution from the $n^\text{th}$ LL as
\begin{align}
\label{div omega init}
    \Omega_\mu^{f=1,\ell=n}
    =-\frac{N_c\left|eB/3\right|}{(2\pi)^2}
    \int_{-\infty}^{+\infty}dk
		\sum_{\xi=\pm1}
		(\mu-E_\xi^n)\theta(\mu-E_\xi^n)
\end{align}
where
\begin{equation}
    E_\xi^n=\sqrt{\big(\sqrt{m^2+k^2}+\xi b\big)^2+2(|eB|/3)n}.
\end{equation}
If $n$ is even, then the integrand of the $f=2,\ell=n/2$ term is identical to that of (\ref{div omega init}), and its contribution will simply be double that of (\ref{div omega init}). These terms have the same sign, so their contributions to the limit in (\ref{curvature limit full}) must add constructively rather than cancel. From here on we will only consider the $f=1,\ell=n$ term. 

Let us consider only the region where $b>\mu_n.$ The $\xi=+1$ term vanishes because $(\sqrt{m^2+k^2}+b)^2\geqslant b^2>\mu_n^2=\mu^2-2(|eB|/3)n$, from which it follows that the $\theta$ function in (\ref{div omega init}) vanishes. Taking the derivative with respect to $m^2$ and then the limit $m\to0$ for the remaining $\xi=-1$ term gives
\begin{align}
\label{div omega final}
    \left.
    \frac{\partial\Omega_\mu^{f=1,\ell=n}}
    {\partial(m^2)}
    \right|_{m=0,b>\mu_n}
    &=\frac{N_c\left|eB/3\right|}{(2\pi)^2}
    \lim_{m\to0}
    2\int_0^\infty dk
	\frac{\partial E_-^n}{\partial(m^2)}
	\theta(\mu-E_-^n)
    \nonumber\\\nonumber\\
    &=\frac{N_c\left|eB/3\right|}{(2\pi)^2}
    \int_{b-\mu_n}^{b+\mu_n}
		\frac{dk}{\sqrt{(k-b)^2+2(|eB|/3)n}}
	    \left(1-\frac bk\right),
\end{align}
where we have used the fact that the argument of the $\theta$ function in the limit $m\to0$, $\mu-\sqrt{(k-b)^2+2(|eB|/3)n}$, is positive if and only if $\mu^2>(k-b)^2+2(|eB|/3)n$, which is equivalent to $\mu_n>|k-b|$. 

The last integral in (\ref{div omega final}) can be written as the difference of two integrals, distributing the rightmost term in parentheses. The first such integral is independent of $b$ and finite: it has the exact solution $2\tanh^{-1}(\mu_n/\mu)$. The second integral can also be solved exactly for the case $b=\mu_n+\epsilon,$ and it can be shown in this way to diverge as $\epsilon\to0$. The divergence is more easily demonstrated, however, by showing that the integral has a divergent lower bound:
\begin{align}
\label{div lower bound}
    \int_{b-\mu_n}^{b+\mu_n}dk
		\frac1{\sqrt{(k-b)^2+2(|eB|/3)n}}\frac bk
    &\geqslant
    \int_{b-\mu_n}^{b+\mu_n}dk
		\frac1{\sqrt{(\mu_n)^2+2(|eB|/3)n}}\frac bk
	\nonumber\\\nonumber\\
	&=\frac{\mu_n+\epsilon}\mu
	\int_\epsilon^{2\mu_n+\epsilon}dk\frac1k.
\end{align}
The final expression in (\ref{div lower bound}) is clearly finite for $\epsilon>0$, but diverges in the limit $\epsilon\to0$. Along with the extra negative sign from (\ref{div omega final}) and the preceding arguments, this proves the original claim. 

An alternative approach is to begin by assuming $T>0$ and calculating the quantity $\partial\Omega/\partial(m^2)$ exactly at $(m,b)=(0,\mu_n).$ As mentioned in Appendix \ref{app:alpha} and shown explicitly in (\ref{eqn:omega mu+T}), the terms $\Omega_\mu$ and $\Omega_T$ can be combined into a single expression. One can then take the derivative of this expression with respect to $m^2$ and obtain an exact result, which is finite for $T>0$ but diverges in the limit $T\to0$. Thus, the negative curvature near $(m,b)=(0,\mu_n)$ (and hence remnant mass) even exists for sufficiently small but nonzero $T$.

\section{Approximation of remnant $\bm{T_c}$}
\label{app:remnant T_c}

In Fig. \ref{fig:critical T}, we have plotted the ``remnant critical temperature," that is, the temperature at which the small remnant mass over the region of intermediate $\mu$ vanishes. As discussed in Sec. \ref{sec:GL at B>0}, the remnant mass (and hence remnant critical temperature) cannot be calculated using the GL coefficients given by (\ref{eqn:coef prefactors})--(\ref{eqn:alpha}). This is because the remnant mass is related to the discretization of Landau levels (see Appendix \ref{app:remnant mass}), whereas the $\alpha$ coefficients in (\ref{eqn:alpha}) are calculated using the Euler-Maclaurin formula, which approximates the Landau sums as integrals. Therefore, a different technique is required to efficiently calculate the remnant critical temperatures.

As shown in Appendix \ref{app:remnant mass}, the remnant mass is related to singularities in the quantity $\partial\Omega/\partial(m^2)|_{m=0,b=\mu_n}$ for each integer $n$ such that $\mu_n=\sqrt{\mu^2-2(|eB|/3)n}$ is real. The singular behavior occurs only at zero temperature; if we let $T>0$ and include the finite-temperature contribution $\Omega_T$, the derivative becomes finite. The idea is to approximate the derivative in the small-$T$ limit, which, as we will see, yields an expansion of the form:

\begin{equation}
\label{eqn:small T expansion}
    \frac{\partial\Omega}{\partial(m^2)}\Bigg|_{m=0,b=\mu_n}=A+B T^2+C\ln T,
\end{equation}

where $A, B$ and $C$ are functions of $eB$, $\mu$, and $n$. It is then easy to show that the quantity in (\ref{eqn:small T expansion}) vanishes at

\begin{equation}
\label{eqn:T_c^n}
    T^{(n)}=\sqrt{\frac C{2B}W\left(\frac{2B}C\exp^{-2A/C}\right)},
\end{equation}

Where $W(z)$ is the product logarithm (or Lambert function). The critical temperature $T_c$ at fixed $eB$ and $\mu$ is then the maximum $T^{(n)}$, where $n$ ranges over all integers for which $\mu_n$ is real.

After taking the derivative with respect to $m^2$ of the LLL terms and letting $m\to0$, the integrals in each expression can be solved analytically, giving the following closed-form solutions,
\begingroup
\allowdisplaybreaks
\begin{align}
\label{eqn:LLL derivatives}
\frac{\partial\Omega_{vac}^{LLL}}{\partial(m^2)}\bigg|_{m=0}&=\frac{N_c\left|e B\right|}{(2\pi)^2}\left[\ln\left(\frac{2b}\Lambda\right)+\frac\gamma2-\left(\frac b\Lambda\right)^2\,{}_2F_2(1,1;3/2,2;-b^2/\Lambda^2)\right]
\\\nonumber\\
\frac{\partial\Omega_\mu^{LLL}}{\partial(m^2)}\bigg|_{m=0}&=\frac{N_c\left|e B\right|}{(2\pi)^2}\ln\left(\frac{|\mu-b|}b\right)
\\\nonumber\\
\frac{\partial\Omega_T^{LLL}}{\partial(m^2)}\bigg|_{m=0}&=\frac{N_c\left|e B\right|}{(2\pi)^2}\left[\operatorname{Re}\psi\left(\frac12+i\frac{|\mu-b|}{2\pi T}\right)-\ln\left(\frac{|\mu-b|}{2\pi T}\right)\right]
\\\nonumber\\
&\approx\frac{N_c\left|e B\right|}{(2\pi)^2}\left[-\frac{\pi^2}{6(\mu-b)^2}T^2\right],
\end{align}
\endgroup
Where ${}_2F_2$ is the generalized hypergeometric series. The corresponding vacuum HLL term can be expressed as a single integral,
\begin{equation}
   \frac{\partial\Omega_{vac}^{HLL}}{\partial(m^2)}\bigg|_{m=0}=\frac{\left|e B\right|}{(2\pi)^2}\int_{1/\Lambda^2}^\infty ds\left(\frac1{e^{(2/3)|eB|s}-1}+\frac2{e^{(4/3)|eB|s}-1}\right)\left(\frac{2b}{\sqrt s}F(b\sqrt s)-\frac1s\right),
\end{equation}
Where $F(z)$ is the Dawson function.

It remains to calculate the terms corresponding to the HLL fermion contribution at zero and finite temperature. First, we observe that the expressions for these two contributions, given by (\ref{eqn:omega mu HLL}) and (\ref{eqn:omega T}), can be combined into one expression,
\begin{equation}
\label{eqn:omega mu+T}
    \Omega^{f,HLL}_{\mu+T} 		=-2\frac{N_c\left|e_f B\right|}{(2\pi)^2}
		\frac{1}{\beta}
		\int_0^\infty dk
			\sum_{\ell\xi\epsilon}\ln{
				\left(
					1+e^{-\beta(|E_\ell|+\xi\mu)}
				\right)}.
\end{equation}
Let us first consider the contribution to the above term from a single Landau level $\ell=n$. Taking the derivative with respect to $m^2$, then letting $m\to0$, and finally integrating by parts, one can show that
\begin{equation}
    \label{eqn:mu+T after IBP}
    \frac{\partial\Omega_{\mu+T}^{f,n}}{\partial(m^2)}\bigg|_{m=0}=\frac{N_c\left|e_f B\right|}{(2\pi)^2}\frac\beta4\int_0^\infty dk\sum_{\epsilon=\pm1}\operatorname{sech}^2\left[\frac\beta2\left(\sqrt{(k+\epsilon b)^2+a_n}-\mu\right)\right]
    \frac{k+\epsilon b}{\sqrt{(k+\epsilon b)^2+a_n}}v(k),
\end{equation}
where $a_n\equiv2|e_fB|n$ and
\begin{equation}
    v(k)=\tanh^{-1}\left(\frac{k+\epsilon b}{\sqrt{(k+\epsilon b)^2+a_n}}\right)-\epsilon\frac b{\sqrt{b^2+a_n}}
    \tanh^{-1}\left(\frac{\epsilon b(k+\epsilon b)+a_n}{\sqrt{b^2+a_n}\sqrt{(k+\epsilon b)^2+a_n}}\right).
\end{equation}
In (\ref{eqn:mu+T after IBP}), we have omitted the $\xi=+1$ terms because, in the large $\beta$ limit, their contributions are exponentially suppressed. We also note that the integration by parts in the previous step involves a delicate cancelation of divergences in the boundary term, resulting in two integrals (corresponding to $\epsilon=\pm1$) that are both individually finite, as can be checked by careful analysis. 

As $\beta\to\infty$, the integrand becomes sharply peaked where the argument of $\operatorname{sech}^2$ vanishes, which always occurs at the values $k=|\mu_n\pm b|$, where $\mu_n=\sqrt{\mu^2-a_n}$. We will refer to the peaks at $k=\mu_n+b$ and $k=|\mu_n-b|$ as the \emph{major} and \emph{minor} peaks, respectively. The major peak always occurs in the $\epsilon=-1$ term, whereas the minor peak splits into three cases: (1) if $\mu_n>b$, the minor peak occurs in the $\epsilon=+1$ term; (2) if $\mu_n<b$, the minor peak occurs in the $\epsilon=-1$ term; (3) if $\mu_n=b$, then the minor peak (which coincides with an integrable singularity at $k=0$) occurs in both the $\epsilon=\pm1$ terms. We will sketch the procedure for expanding the integral over the major peak in powers of $T$ since the method applied to the minor peak (in the singular and non-singular cases) is similar. 

The major peak always occurs in the $\epsilon=-1$ term of the integrand. We proceed in three steps. First, we transform the integral using $x=\sqrt{(k-b)^2+a_n}-\mu$. Although this transformation is not globally monotonic, it is monotonic over a neighborhood of the point $k=\mu_n+b$ of some radius $\delta$, and $\delta$ will become large compared to the peak width at sufficiently small $T$. Regions far from the peak are exponentially suppressed by the $\operatorname{sech}^2$ term, so we can approximate the transformed integral as an integral over the entire real line, giving
\begin{align}
    I_\text{major}&\equiv\frac\beta4\int_{\mu_n+b-\delta}^{\mu_n+b+\delta} dk\operatorname{sech}^2\left[\frac\beta2\left(\sqrt{(k-b)^2+a_n}-\mu\right)\right]\frac{k-b}{\sqrt{(k-b)^2+a_n}}v(k)\\\nonumber\\
    &\approx\frac\beta4\int_{-\infty}^{+\infty} dx\operatorname{sech}^2\left[\frac\beta2x\right]v\left(b+\sqrt{(x+\mu)^2-a_n}\right)
\end{align}
The $\operatorname{sech}^2$ peak is now centered at $x=0$. Because it becomes very narrow as $\beta\to\infty$, we are justified in expanding the other term in the integrand, $v(\cdots)$, about $x=0$ and then integrating term by term. We have
\begin{align}
    v\left(b+\sqrt{(x+\mu)^2-a_n}\right)&=\tanh^{-1}\left(\frac{\sqrt{(x+\mu)^2-a_n}}{x+\mu}\right)+\frac b{\sqrt{b^2+a_n}}\tanh^{-1}\left(\frac{-b\sqrt{(x+\mu)^2-a_n}+a_n}{\sqrt{b^2+a_n}(x+\mu)}\right)\\\nonumber\\
    \label{eqn:v(x) expansion}
    &=c_0+c_1x+c_2x^2+O(x^3)
\end{align}
where
\begin{align}
    c_0&=\tanh^{-1}\left(\frac{\mu_n}\mu\right)+\frac b{\sqrt{b^2+a_n}}\tanh^{-1}\left(\frac{-b\mu_n+a_n}{\sqrt{b^2+a_n}\mu}\right)\\\nonumber\\
    c_2&=-\frac\mu{2\mu_n(b+\mu_n)^2}.
\end{align}
(We omit $c_1$ because all terms with odd powers of $x$ vanish upon evaluating the integral.) Finally, we make the transformation $y=(\beta/2)x$ and integrate term by term, using the formula $\int_{-\infty}^{+\infty}y^n \operatorname{sech}^2y\,dy=(-1)^{n/2}(2^{2-n}-2)\pi^nB_n$, which gives
\begin{equation}
    I_\text{major}=\left[\tanh^{-1}\left(\frac{\mu_n}\mu\right)+\frac b{\sqrt{b^2+a_n}}\tanh^{-1}\left(\frac{-b\mu_n+a_n}{\sqrt{b^2+a_n}\mu}\right)\right]-\left[\frac{\pi^2}{6}\frac1{(\mu_n+b)^2}\frac\mu{\mu_n}\right]T^2+O(T^4).
\end{equation}
A similar approach can be applied to the minor peak, with the only added subtlety being that in the singular case $\mu_n=b$, the peak of the integrand lies on an integrable singularity at $k=0$. In this case, the expansion of $v$ given by (\ref{eqn:v(x) expansion}) will include a logarithmic term $c_\text{sing}\ln(x)$, which can be integrated using the formula $\int_0^\infty\ln(y)\operatorname{sech}^2(y)dy=\ln(\pi/2)-\gamma$. The results are
\begin{align}
    I_\text{minor}^\text{nonsing}&=\left[\tanh^{-1}\left(\frac{\mu_n}\mu\right)-\frac b{\sqrt{b^2+a_n}}\tanh^{-1}\left(\frac{b\mu_n+a_n}{\sqrt{b^2+a_n}\mu}\right)\right]-\left[\frac{\pi^2}{6}\frac1{(\mu_n-b)^2}\frac\mu{\mu_n}\right]T^2+O(T^4)\\\nonumber\\
    I_\text{minor}^\text{sing}&=\left\{\tanh^{-1}\left(\frac{\mu_n}\mu\right)+\frac{\mu_n}\mu\left[\ln\left(\frac{\pi\sqrt{a_n}}{4\mu\mu_n}\right)-\gamma\right]\right\}+\left\{\frac{\pi^2}{24}\frac{a_n(a_n-2\mu^2)}{\mu^3\mu_n^3}\right\}T^2+\frac{\mu_n}\mu\ln T+O(T^4).
\end{align}
With these expressions at hand, one can efficiently compute the coefficients $A,B$ and $C$ of (\ref{eqn:small T expansion}), and then it is straightforward to determine $T_c$ as explained above.

\end{document}